\documentclass[11pt]{article}
\usepackage[utf8]{inputenc}
\usepackage{amsbsy,amsfonts,amsmath,amssymb,amsthm
,bbm,enumerate,euscript,graphicx,color}
\usepackage{subcaption,lipsum}
\usepackage{hyperref}

\theoremstyle{plain}

\theoremstyle{definition}

\theoremstyle{remark}

\newcommand{\R}{\mathbb{R}}
\newcommand{\N}{\mathbb{N}}

\newcommand{\so}{\mathrm{SO}(3)}
\numberwithin{equation}{section}
\DeclareMathOperator*{\argmax}{arg\,max}

\title{\bf {Random marked nested tessellations applied to the modelling of deformation twinning in polycrystalline materials}}

\author{Oleksandr Kornij\v{c}uk\textsuperscript{a},
Luděk Heller\textsuperscript{b}, Zbyněk Pawlas\textsuperscript{a}, Viktor Bene\v{s}\textsuperscript{a}}

\begin{document}

\maketitle

\thanks{\noindent\textsuperscript{a}Department of Probability and Mathematical Statistics, Faculty of Mathematics and Physics, Charles University, Sokolovsk\'{a} 83, 18675 Praha 8, Czech Republic; \textsuperscript{b}Department of Functional Materials, Institute of Physics of the Czech Academy of Sciences, Na Slovance 2, 18221 Praha 8, Czech Republic}

\begin{abstract}
Stochastic geometry provides a powerful framework for modelling complex random structures, with applications in physics, materials science, biology, and other fields. The three-dimensional microstructure of polycrystalline materials is usually modeled by a randomly marked tessellation, where the marks correspond to crystallographic orientations. The purpose of this study is to extend the modelling approach to a finer scale, focusing on the subcells that emerge when a material specimen is exposed to mechanical loading. Specifically, the deformation twinning gives rise to nested tessellation, where the subcells are parallel twin lamellae and their complement is embedded within the original mother cells. The aim of this study is to develop a parametric mathematical model of marked nested tessellation and to realize it using stochastic simulations. We were able to deal with this model using computational tools. The sensitivity of the model to selected key parameters was investigated using statistical methods. As an application, a numerical simulation of the stress and strain fields resulting from deformation twinning is provided, and the contribution of the subcells to the total strain energy density under varying initial conditions was evaluated.
This study highlights the dynamic capabilities of stochastic geometry in modelling a phenomenon that changes the microstructure.\end{abstract}

\vspace{0.2cm}

\noindent
\textit{Keywords.} deformation twinning, nested tessellation, Schmid factor, stochastic simulation, total strain energy density

\vspace{0.5cm}
\noindent
\textit{AMS Classification}: 60D05, 62P35

\section{Introduction} \label{sec:intro}
Tessellation \cite{Okabe} is a geometrical structure comprising disjoint space-filling cells. A particularly relevant subclass is that of Laguerre tessellations (called power diagrams in \cite{Aurenhammer}), which serve as suitable models for the microstructure of polycrystalline materials in $\R^3$; see \cite{Petrich}. Due to the natural variability in such microstructures, stochastic models are commonly employed, that is, random tessellations \cite{Chiu}, particularly random Laguerre tessellations \cite{Lautensack}.

In stochastic geometry \cite{Chiu}, marking is a standard operation. For example, for a random point process in the Euclidean space, assigning a random mark (from an arbitrary space) to each point leads to a marked point process. In this study, we deal with marked tessellations, in which we assign a mark to each cell. In materials science applications, \cite{Pawlas, Karafiatova} used marks to describe cubic crystallographic orientations, represented as elements of the rotation group $\so$ \cite[Chapter 3]{Morawiec}. Here, we extend the marking approach by incorporating additional physical features of crystal defects \cite{Kelly}. Another important operation on tessellations is nesting, which leads to so-called nested tessellations \cite{Schreiber}, where each cell is further subdivided. We refer to the original cells as mother cells and the newly created cells as subcells.

The aim of the present study is to introduce a system of marks for a Laguerre tessellation with cubic orientations and consequently establish a random nested tessellation that serves as a parametric microstructural model for deformation twinning \cite[Chapter 11]{Kelly}. There are two types of subcells: twinning lamellae (defects) and interlamelar spaces (matrices). This nested tessellation is no longer Laguerre, it is not analytically tractable, and we must rely on stochastic simulation and numerical computing. 

The model was applied to simulate the internal stresses from deformation twinning in B2 cubic austenite in near-equiatomic NiTi \cite{Otsuka}. In fact, these deformation twins found in the cubic austenite of NiTi are relicts of the stress-induced martensitic transformation and the so-called kwinking process activated in the B19' monoclinic martensite phase \cite{Seiner}. The kwinking process in martensite is a plastic deformation process that forms bands or reoriented crystal structures such that the (20$\bar{1}$) plane (in the notation of the Miller indices \cite[Section 2.9]{Morawiec}) becomes a mirror plane of the initial crystal and the reoriented crystal \cite{Molnarova2021,Molnarova2023}. After reverse transformation, the (20$\bar{1}$) martensite deformation twins are transformed into one of the \{114\} deformation twins of austenite as identified by experimental observations \cite{Chen2019,Chen201902,Alacron}. Marked nested tessellation was used to model \{114\} deformation twins. Numerical finite element analysis was used to compute the internal stresses resulting from the strains associated with deformation twins. Although the computation provides the stress and strain tensor fields in a bounded specimen, we restrict ourselves to a scalar total strain energy density and perform a regression analysis of this dependent variable with model parameters as independent variables, namely the texture (property of the orientation distribution) and macroscopic strain, which is proportional to the volume fraction of twinning lamellae in the mother cells.

The paper begins with a background on random crystallographic orientations and marked Laguerre tessellations. Then, we summarize an introduction to the physical context of the problem, namely, deformation twinning based on the Schmid factor and the total strain energy density. 
Section \ref{sec:results} presents the main results of the study, where a parametric model of marked nested tessellation with twin lamellae was introduced and investigated. A Python-based computer program was developed and used in conjunction with Neper software \cite{Quey} to perform a stochastic simulation of the model followed by a numerical simulation using the MSC Marc solver \cite{MSCA}. A statistical evaluation of the sensitivity of the model with respect to the input parameters is presented in Section \ref{sec:Numerical}.

\section{Mathematical background}
We deal with the Euclidean space $\mathbb{R}^d$ equipped with a Borel $\sigma$-algebra ${\cal B}^d$, mostly $d=3$ because the modelling of the 3D material microstructure is of interest. We denote the Lebesgue measure of the set $A\in{\cal B}^3$ as $|A|$. For $u,v\in\R^3$, their vector, tensor and inner product are written as $u\times v$, $u\otimes v$, $\langle u,v\rangle$, respectively. The notation $\|\cdot\|$ stands for the Euclidean norm. We work with the column vectors $u$, where $u^T$ denotes their transpose.
\subsection{The space $\so$}
The main component of our modelling and analysis was the crystallographic orientation of the cubic lattice \cite[Section 6.2]{Morawiec}. The group of all rotation matrices in $\mathbb{R}^3$ is defined as
   $$     \so = \{ G \in \R^{3 \times 3} : \, G G^T = \mathbb{I}_3, \, \det G = 1 \} .$$ 		
    Let $\mathcal{O}$ be the subgroup of $\so$ given by the rotational symmetry of a cube, it has $24$ elements. We consider two elements $G_1, G_2\in \so$ to be equivalent if there exists $R \in \mathcal{O}$ such that $G_1 = R G_2$.
 Thus, we consider the quotient space $\so / \mathcal{O}$. If $G_1, G_2$ are not equivalent, their disorientation angle is
\begin{equation} \label{eq:dis}
    \textnormal{dis} (G_1, G_2 ) = \min_{R \in \mathcal{O}} \arccos \frac{ \textnormal{tr} (G_1^{-1} R G_2 ) -1 }{2}, \quad G_1, G_2 \in \so,
\end{equation}
where $\textnormal{tr}$ is the trace of the matrix.
Moreover, we can characterize a rotation by means of the way it acts in one direction with respect to another. 
    Let $v, u \in \mathbb{R}^3$ be nonzero vectors and $G \in \so$. Then we call
\begin{equation*} 
    t(G, v, u) = \max_{R \in \mathcal{O}} \frac{v^T RG u}{\| v \|\, \| u \|}
\end{equation*}
    the tilt of $G$ with respect to directions $v$ and $u$. 
    Suppose that $u$ is a direction in the coordinate system of $\R^3$ and $v$ is a direction with respect to the coordinate system of the cubic lattice. Then $t(G, v, u)$ computes the cosine of the angle between $u$ in all equivalent coordinate systems of the cube and the direction $v$, and selects the largest value.

\begin{figure}[t]
\centering
\subfloat[$\kappa=0$]{%
\resizebox*{6cm}{!}{\includegraphics{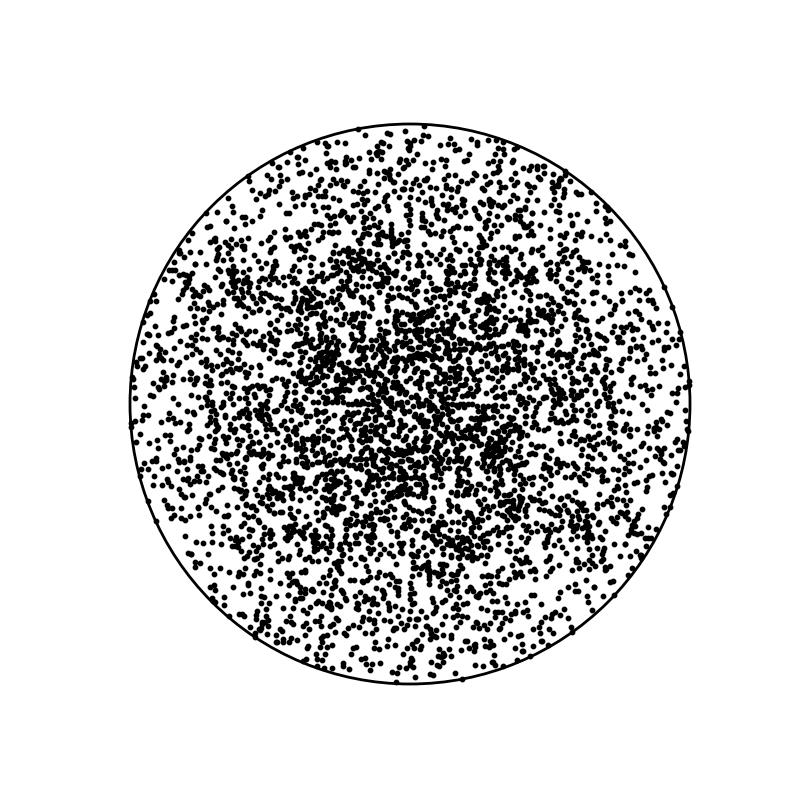}}}\hspace{10pt}
\subfloat[$\kappa=30$]{%
\resizebox*{6cm}{!}{\includegraphics{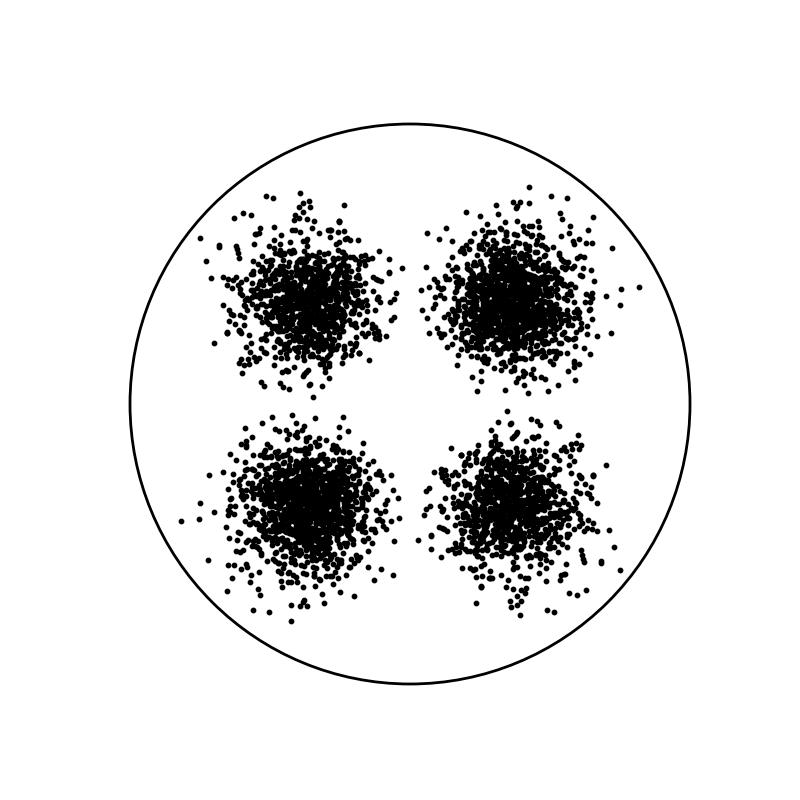}}}
\caption{Inverse pole figures with respect to the direction $z=(0,0,1)^T$ with simulated orientations  from the density \eqref{Preferential function} for $u=(0,0,1)^T,\; v=(1,1,1)^T$ and two different parameters $\kappa$. The centres of four clusters in (b) correspond to preferred orientations $\langle 111\rangle$.} \label{ipfs}
\end{figure}

Next, we introduce random orientation as a random variable $X$ with values in $\so$.
According to \cite[Chapter 11]{Halmos}, there exists a unique Haar measure $\mu$ in the Borel subsets of $\so$ such that $\mu(\so)=1$. A nonnegative Borel measurable function $f$ in $\so$ is called the probability density of orientations (shortly 'odf') if
$$\int_{\so}f(G)\,\mu({\rm d}G)=1.$$ We can say that $X$ has a distribution with odf $f$. In particular, if $f=1$ is constant on $\so$, then $X$ has a uniform distribution, and we say that there is no (crystallographic) texture. We consider a parametric class of odf's depending on the tilt $t$:
\begin{equation} \label{Preferential function}
        f (G) \propto \exp ( \kappa  t(G, v, u)), \ G \in \so,
    \end{equation} 		
where $u, v \in \mathbb{R}^3$ are nonzero vectors and $\kappa \geq 0$. Note that $f(G) = f(RG)$ for all $R \in \mathcal{O}$ and thus $f$ can also be viewed as a density on $\so/\mathcal{O}$. There is a preferred orientation because with increasing $\kappa$ the alignment between $v$ and $Gu$ is more probable, which leads to a stronger texture. In Figure \ref{ipfs} the inverse pole figures \cite[Section 10.3]{Morawiec} of the simulated orientations are presented for $\kappa=0$ (uniform) and $\kappa=30$, with respect to the direction $z=(0,0,1)^T$.

\subsection{Marked Laguerre tessellation in $\R^3$} \label{sec:marked}
Let ${\cal I}$ be an index set that is at most countable with at least two elements. A tessellation of space $\mathbb{R}^3$ (also called a 3D tessellation) is a locally finite system $T=\{C_i,\,i\in{\cal I}\}$ of closed space-filling cells $C_i$ that have pairwise disjoint interiors \cite{Okabe}. Laguerre tessellation is a special class of tessellations based on a set of weighted generators $\{(x_i, w_i), i \in{\cal I} \}$, where $x_i\in\mathbb{R}^3$ are distinct points (generators) and $w_i\in\mathbb{R}$ are weights.     
The cells of Laguerre tessellation are given by
    \begin{equation*}
        C_i = \{ x \in \mathbb{R}^3 :   \, \| x - x_i \|^2 - w_i \leq \| x - x_j \|^2 - w_j, \, \forall j \in {\cal I}\}, \quad i \in {\cal I},
    \end{equation*}
    they are convex polyhedra. We say that two cells are neighbors ($C_i\sim C_j$) if they share a common face.
Note that not all generators necessarily form nonempty cells. We denote by $T_L = \{ C_i \neq \emptyset : i \in {\cal I} \}$ a Laguerre tessellation with a given set of generators and weights. Laguerre tessellations are face-to-face and normal, see \cite{Aurenhammer} for details and further properties.

Let $\Phi$ be a stationary random marked point process (see \cite{Chiu}) with points in $\mathbb{R}^3$ and real marks. Then $\Phi$ induces a random Laguerre tessellation in $\mathbb{R}^3$ with the points being generators and the weights corresponding to the marks (see \cite{Lautensack}).

Next, consider a random Laguerre tessellation in a compact convex domain $Q\subseteq\mathbb{R}^3$ in the sense that $Q$ is an observation window of a larger tessellation. To create such a tessellation we will use plus sampling, that is, we consider a marked point process $\Phi$ in a larger window $Q_e$, $Q\subseteq Q_e$, such that all cells of the corresponding Laguerre tessellation hitting $Q$ are completely contained in $Q_e$ (with a probability close to 1). 

Another aim is to achieve Laguerre tessellation with a prescribed distribution $\cal V$ of cell volumes. For a given number $n\in\N$ and a set $\{x_1,\dots,x_n\}$ of generators in $Q_e$ each of which will form a cell, and independent identically distributed random variables (iid) $X_i,i=1,\dots,n,$ with the distribution $\cal V$, there exists \cite{Bourne} a set of weights $w_i,i=1,\dots,n$ such that $\{(x_i,w_i), i=1,\dots,n\}$ forms a Laguerre tessellation in $Q_e$ with $n$ cells of volumes \begin{equation}\label{vol}V_i=|Q_e|\frac{X_i}{\sum_{j=1}^nX_j}, \quad i=1,\dots,n.\end{equation} An effective algorithm to calculate the weights $w_i$ is available; see \cite{Bourne}.

Furthermore, we consider a random marked Laguerre tessellation \cite{Pawlas} in which the mark space is $\so/{\cal O}$, that is, the marks represent equivalence classes of crystallographic orientations. The realization of a random marked Laguerre tessellation consists of pairs $(C_i,G_i)$, $i\in{\cal I}$, of cells and orientations. An independent marking (IM) is defined so that the marks $G_i$ are iid random elements in $\so/{\cal O}$, which are independent of the random Laguerre tessellation.

We also deal with a particular case of dependent marking called moving average (MA) \cite{Pawlas}. Let $\{(C_i,G_i),\,i\in{\cal I}\}$ be an independently marked Laguerre tessellation with uniform orientation distribution. Each orientation $G_i$ is represented by a unit quaternion $q_i$, $i\in{\cal I}$, selected from the fixed asymmetric domain of the quaternion representation \cite{Morawiec}.
We set \begin{equation}\label{ma}   
        \tau_i = \frac{\sum_{j \in{\cal I}} \mathbb{I} \{C_i \sim C_j\} q_j}{\left\| \sum_{j \in {\cal I}} \mathbb{I} \{C_i \sim C_j\} q_j \right\|}, \quad i\in{\cal I},\end{equation}
        where scalar multiplication and addition are interpreted component-wise. Transform quaternions $\tau_i$ back into rotation matrices $\bar{G}_i$, then $\{(C_i,\bar{G}_i),\,i\in{\cal I}\}$ form the MA model.

\section{Physical models}
In this section, we present some reasoning from crystallography \cite{Kelly} and materials physics \cite{Sadd} that yields the motivation and background for our stochastic modelling. Deformation twinning is a phenomenon described in Section \ref{sec:dt}, where we consider a single crystal and assume that its coordinate system $(x,y,z)$ coincides with that of $\R^3$. The crystal possesses a cubic lattice orientation $G\in\so$, which defines another coordinate system $(x',y',z')$ given by the edges of the cubic lattice. These systems are related through $$ (x', y', z')^T = G\, (x, y, z)^T.$$ Section \ref{sec:stress}, on the other hand, considers a polycrystal, which in the context of mathematical modelling corresponds to a tessellation (analogously, a crystal corresponds to a single cell). The tensorial internal stress field in the polycrystal after deformation twinning is mentioned, and the scalar total strain energy density is defined. 

\subsection{Deformation twinning} \label{sec:dt}
\begin{figure}[t]
\centering
\includegraphics[width=\linewidth]{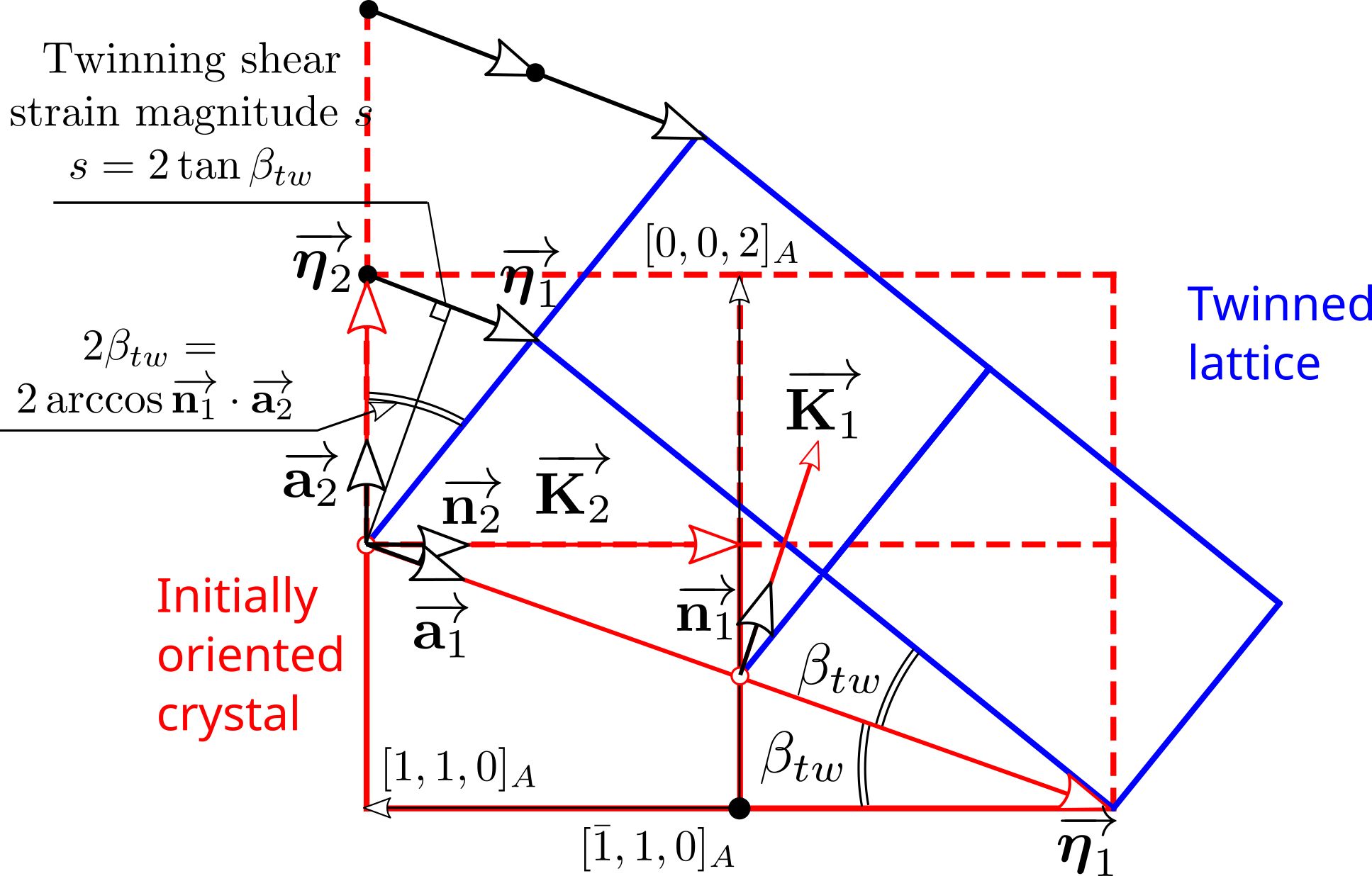}
\caption{Cubic lattice projected onto $(\bar{1}10)$ plane in the configuration before (red) and after (blue) twinning described the twinning elements $K_1=(\bar{1}\bar{1}4),K_2=(\bar{1}\bar{1}0),{\eta}_1=[\bar{2}\bar{2}\bar{1}],{\eta}_2=[001]$. Index $A$ means austenite, cf. Section \ref{sec:intro}.}
\label{fig01}
\end{figure}

Assume in the following fixed loading direction $D_l\in\mathbb{R}^3$ and crystal $C$ with orientation $G\in\so$. According to \cite[Section 11.2]{Kelly}, deformation twinning is described in the crystal coordinate system by a twinning plane $K_1$ with unit normal vector $n_1$, a twinning direction $\eta_1$ with unit vector $a_1$, and a conjugate plane $K_2$ with unit normal vector $n_2$ and a direction $\eta_2$ with unit vector $a_2$. The twinning consists of the rotation of the crystal lattice around the vector $n_1 \times a_1$ by an angle $2 \beta_{tw}$; see Figure \ref{fig01} for a special choice of twinning parameters.
The deformed regions in $C$ that have taken up a twin orientation are called lamellae.
Let $R_t = 2 a_1 \otimes a_1 - I_3$ be an orientation matrix, then
$$\beta_{tw} =\arccos ( \langle a_2 , n_1 \rangle ).$$ The product $GD_l$ rotates $D_l$ to the crystal coordinate system.
The Schmid factor $$\chi (G,a_1,n_1)= \langle n_1 , GD_l \rangle \, \langle a_1 , GD_l \rangle$$ 
takes values in the interval $-0.5\leq \chi (G,a_1,n_1)\leq 0.5$ and allows the introduction of the propensity for twinning of a marked crystal $(C,G),\; G\in \so$, as
\begin{equation}\label{prope}\;    \Psi (G, a_1 , n_1) = \max_{R\in{\cal O}}\chi(G,R a_1,Rn_1).\end{equation}
We assume that  \begin{equation*}
        \overline{R} (G) = \argmax_{R \in \mathcal{O}} \chi (G,Ra_1,Rn_1) 
\end{equation*} is unique. Set
$\overline{n}_1=\overline{R}(G)n_1$, $\overline{a}_1=\overline{R}(G)a_1$, ${s}=2\tan(\arccos\langle\overline{n}_1,\overline{a}_1\rangle),$ then the twinning normal vector expressed in $\R^3$ is 
\begin{equation}\label{E1.33}\vec{n}(G)=G^{-1}\overline{n}_1.\end{equation}
The twinning reorients the twin lamellae from their original orientation $G$ to $\overline{R}_t(G)G$, where \begin{equation}\overline{R}_t(G)=2\overline{a}_1\otimes\overline{a}_1-I_3.\label{reor}\end{equation} 

Having defined the propensity for twinning, we next quantify the strain induced by twinning,
which measures the relative change in vectors after twinning with respect to their initial state before twinning. It is a symmetric tensor of rank two calculated from the deformation gradient that maps the lattice vectors prior to twinning to their state after twinning.

For a fixed loading direction $D_l$, the deformation gradient $\overline{F}$ is a matrix calculated in the crystal coordinate system as follows:
$$    \overline{F} = I_3 + {s}\,\overline{a}_1\otimes \overline{n}_1,
$$
and the Lagrangian finite-strain tensor $\overline{E_L}$ for a crystal with orientation $G \in \so$ is defined as
\begin{equation}\label{strcond}
        \overline{E_L} = \frac{1}{2} \big(\overline{F}^T \overline{F} - I_3\big) .
\end{equation}
\noindent
The strain induced by twinning $\overline{E}$, along the uniaxial loading direction $D_l,$ is then calculated from $\overline{E_L}$ as
   $$
        \overline{E} = \big(GD_l\big)^T \overline{E_L} \big(GD_l\big). 
$$
Having defined $\overline{E}$, we considered the volume fraction $V_t$ of the crystal part with deformation twinning. To achieve a macroscopic strain $\varepsilon_m \in (0, 1)$, $V_t$ is calculated based on the assumption of direct proportionality between the macroscopic strain and the strain induced by twinning:
\begin{equation}\label{mstrain}       V_t = \frac{\varepsilon_m}{\overline{E}}.
\end{equation}

\subsection{Internal stress field and strain energy density} \label{sec:stress}
The internal stresses due to the appearance of twin deformation in the form of lamellae were predicted using numerical simulations based on finite elements applied to polycrystal models described by nested tessellations. The algorithm that introduces lamellae considers uniaxial loading, crystal orientations, and Schmid factors. The material behavior considers anisotropic elasticity and isotropic von Mises plasticity with hardening \cite{Sadd}. The internal stress due to deformation twinning was simulated using the structural boundary conditions of the inherent strain implemented in the MSC Marc software \cite{MSCA} for finite element analysis. This condition corresponds to $\overline{E_L}$ \eqref{strcond} in a crystal with orientation $G$.


The internal stresses are evaluated using a strain energy density scalar function $W$ as
\begin{equation*}{\displaystyle W(x)={\frac {1}{2}}\sum _{i=1}^{3}\sum _{j=1}^{3}\sigma _{ij}(x)\varepsilon_{ij}(x),\quad x\in Q},\end{equation*}
where $\sigma _{ij}(x)$, $\varepsilon_{ij}(x)$ are functions of the components of the stress and total strain tensor, respectively. The total strain tensor consists of elastic and plastic parts.
The total strain energy density (TSED) $W$ is a scalar quantity given by
\begin{equation}W=\int_Q W(x)\,{\rm d}x\label{Tsed}.\end{equation}

\section{Results} \label{sec:results}
The location and size of twinning lamellae within cells during deformation twinning exhibit an inherent randomness. To capture this behavior, we developed a parametric model of a marked nested tessellation based partly on the theoretical framework from Section \ref{sec:dt} and partly on some empirical rules from materials physics experience for tessellating individual mother cells by twin lamellae. Random lamellar systems were modeled independently for each cell. Some cells do not contain lamellae a priori because of their low propensity for twinning. In Section \ref{sec:lamel}, we present a method called lamellar growth for simulating twin lamellae, and briefly discuss an alternative method. In Section \ref{sec:stoch} we describe the details of the steps of the entire simulation procedure, both stochastic and numerical, from the underlying model to practical applications.

\subsection{Laguerre tessellation with lamellae} \label{sec:lamel}
\begin{figure}
\centering
\includegraphics[width=11cm]{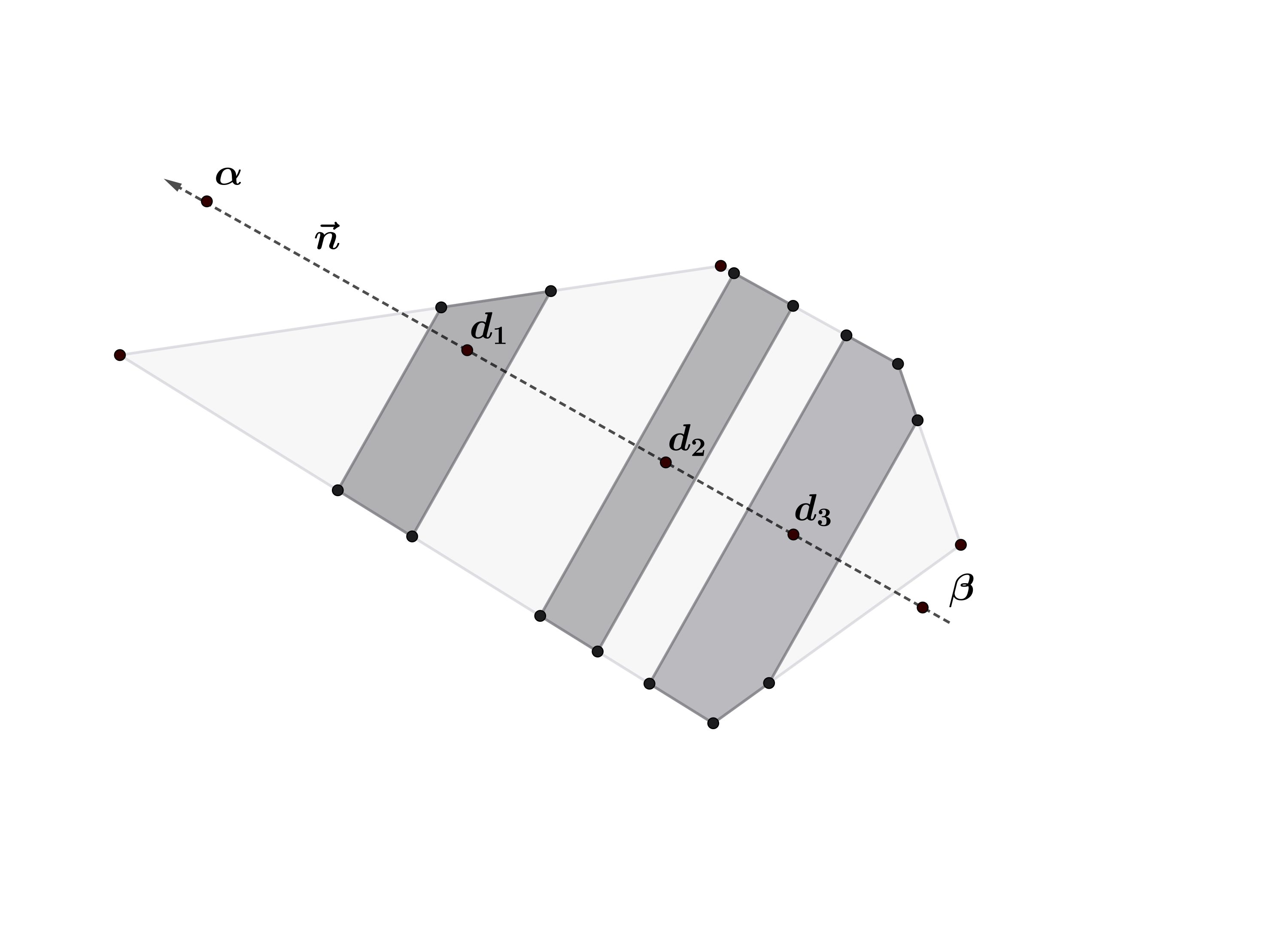}
\caption{Planar projection of a 3D cell with Feret segment $[\alpha ,\beta]$ along the direction $\vec{n}$ and simulated lamellae (gray) of centers $d_i$ and some semi-widths $w_i$. }
\label{fig8}
\end{figure}

In this section, cells of a 3D Laguerre tessellation in the larger window $Q_e$ (cf. Section \ref{sec:marked}) will be divided by lamellae that form parallel slabs within the cells (see Figure \ref{fig8}). Let $C$ be a fixed cell and $\vec{n}\in {\mathbb{R}^3}$ be a normal unit vector from \eqref{E1.33}.
The orthogonal projection of $C$ onto a line defined by $\vec{n}$ and centroid $c$ of $C$ forms line segment $[a,b]$ in $\mathbb{R}^3$. We consider an interval $[\alpha, \beta ] \subseteq \mathbb{R}$ such that for each
$y \in [a, b]$ there exists $t \in [\alpha, \beta]$ such that $y = c + t \vec{n}$. Then $a = c + \alpha \vec{n}$ and $b = c + \beta \vec{n}$. For each $d \in (\alpha, \beta)$ and $w > 0$ such that $(d - w, d + w) \subseteq (\alpha, \beta)$, a lamella $L(d,w)$ with center $d$ and width $2w$ is defined as
$$ L(d, w) = C \cap K_{a}(d + w) \cap K_{b}(d - w),$$
    where $K_{a}(t)$ is the closed halfspace in $\mathbb{R}^3$ containing $a$ and bounded by a plane orthogonal to $\vec{n}$ that passes through $c + t \vec{n}$ for $t \in (\alpha, \beta)$. Similarly, $K_{b}(t)$ is defined with respect to $b$.

 The volume function \(V \colon [\alpha, \beta] \to [0, \infty)\) of the cell $C$ is defined as \[
        V(t) = \left|K_{a}(t) \cap C\right|, \quad t \in [\alpha, \beta],
    \]
\( V \) is continuous and increasing, it satisfies \( V(\alpha) = 0 \) and \( V(\beta) = |C| \). The volume of a lamella \( L(d, w) \) within \( C \) is given by
    \[
        | L(d, w)| = V(d+w) - V(d-w), \quad d \in (\alpha+w, \beta-w).
    \]

Let $l_{max}\in\N$ be a fixed parameter that represents the maximum number of lamellae in a cell. Let cell $C$  contain a number $m \leq l_{max}$ of lamellae that together form a lamellar system of $C$. The number $m$ is generated from the initial probability distribution. The Feret diameter of $C$ in direction $\vec{n}$ is equal to $\rho = \|b - a\| = \beta-\alpha$. The lamellar system $\{L(d_i,w_i),\,i=1,\dots ,m\},\;d_1<\dots<d_m$, must meet the following conditions:
\begin{enumerate}[(i)]
    \item lamellae are at least $\rho\gamma$ apart, i.e., $d_j<d_i\implies d_i-w_i-(d_j+w_j)\geq\rho\gamma$, $0<\gamma <1, $
    \item spacing between $L(d_1,w_1)$ and $\alpha$, 
 as well as between $L(d_m,w_m)$ and $\beta$ is no less than $\rho\xi$, $0<\xi<1$,
    \item the semi-widths of lamellae lie in the interval $(\rho\zeta_1 ,\rho \zeta_2 )$, $0<\zeta_1<\zeta_2 \leq 0.5$, and
    \item sum of volumes of lamellae is $\sum_{j=1}^m|L(d_j,w_j)|=V_t |C|$, $V_t\in (0,1)$.
\end{enumerate}
 The method of generation, called lamellar growth, assigns a growth rate $\theta_j\in[1,2)$, $j=1,\dots, m$, to each center $d_j$, which governs its relative contribution to the overall lamellar expansion.
Denote $$\delta_w=\min \left( \frac{d_1 - \alpha - \xi \rho}{\theta_1}, \, \frac{\beta - \xi \rho - d_m}{\theta_m}, \, \min_{j=1, \dots, m-1} \frac{d_{j+1} - d_j - \gamma \rho}{\theta_j + \theta_{j+1}} \right).$$
Then for \( w \in (\zeta_1 \rho, \zeta_2 \rho) \) satisfying
$w\leq \delta_w$ the spacing conditions (i)--(iii) are fulfilled.

Based on this, we define the function \( \Upsilon \) by
\begin{equation*} 
    \Upsilon(w) = \sum_{j=1}^m |L(d_j, \theta_j w)|, \quad  w\in I=(\zeta_1\rho,\,\min(\delta_w,\,\zeta_2\rho)).
\end{equation*}
It can be easily shown that the function $\Upsilon$ is increasing and continuous on $I,$
thus if
\begin{equation}        \min_{w \in I} \Upsilon(w) \leq V_t |C| \leq \max_{w \in I} \Upsilon(w), \label{inta}
\end{equation}
    then, by the intermediate value theorem, there exists a \(w_0 \in I\) such that:
    \begin{equation}
        \Upsilon(w_0) =V_t |C|,\label{upsa}
    \end{equation} thereby satisfying the volume constraint specified in condition (iv).
In the simulation algorithm, given $m$, the candidate center points $d_i,\,i=1,\dots ,m$, are sampled along the interval $(\alpha + \xi \rho + \zeta_1 \rho, \, \beta - \xi \rho - \zeta_1 \rho)$ using the random sequential adsorption method \cite[Section 6.5.3]{Chiu}. We know the quantity $V_t$ from \eqref{mstrain}; therefore condition \eqref{inta} must be verified. If it is satisfied, we can find the solution $w_0$ of Eq. \eqref{upsa} by a grid search, then $$L(d_1,\,w_0\theta_1),\dots ,L(d_m,\,w_0\theta_m)$$ is the output lamellar system for cell $C$. Otherwise, a new number $m$ is generated from the initial distribution, center sampling and lamellar growth is repeated until condition \eqref{inta} is satisfied.

Another method based on simulated annealing \cite{Delahaye} is briefly described. For fixed parameters $\gamma , \xi , \zeta_1, \zeta_2$ and a given $m$, consider the state space $$ S = \{  ( d_1, w_1), \dots, (d_m, w_m)  \text{ satisfying conditions (i)--(iii) above}\}.  $$ 
 Using the optimization method of simulated annealing, we minimize the function
\begin{equation*}
H(s) = \bigg| \sum_{j=1}^m |L (d_j , w_j )| - V_t |C| \bigg|, \quad s \in S.\end{equation*} Thus, condition (iv) is approximated and we obtain a solution $\{L(d_i,w_i)\subseteq C,\,i=1,\dots ,m\}$.

 By the generation of lamellae within some cells of a Laguerre tessellation $T_L$, we obtain a nested tessellation $T_N,$ cf. \cite{Schreiber}, where the mother cells are those of the original tessellation $T_L,$ and the subcells are formed by lamellae and interlamellar spaces (parts of the matrix). Obviously, $T_N$ is not a Laguerre tessellation, because the presence of a single lamella violates the face-to-face condition \cite{Aurenhammer}. 

\subsection{Stochastic and numerical simulation} \label{sec:stoch}
The simulation of nested tessellation that represents the microstructure of a virtual material after deformation twinning and its subsequent numerical and statistical evaluation proceeds in the following steps:
\begin{itemize}
\item{3D random Laguerre tessellation,}
\item{marking of cells by random cubic orientations,}
\item{conditions on the presence of twinning in cells,}
\item{simulation of twin lamellae,}
\item{meshing of nested tessellation,}
\item{numerical simulation of stress and strain fields,}
\item{statistical analysis of total strain energy density.}
\end{itemize}
We shall describe some more details of these steps.
\subsubsection{3D Laguerre tessellation} \label{sec:4.2.1}
Our basic window in which the virtual material is simulated is, without loss of generality, the unit cube $Q=[0,1]^3.$ To account for edge effects, we need to start with an enlarged window $Q_e=[-\Delta, 1+\Delta]^3$, $\Delta>0$, see the discussion on plus sampling in Section \ref{sec:marked}. First, a homogeneous Poisson point process of generators $x_i$, $i=1,\dots,n$, was simulated in $Q_e$ with intensity $\lambda_0$ which corresponds to the mean number of points in $Q$. We then applied the prescribed cell volume distribution of the Laguerre tessellation. In \cite[Fig. 2b]{Heller} it is shown that the distribution of equivalent volume sphere diameters $d^e$ of crystals for the material of interest, a superelastic NiTi alloy, is well approximated by the Gaussian distribution with mean $\mu=5.1$ and standard deviation $\sigma=1.3$ ($\mu$m). Thus, we simulate $d^e_1,\dots ,d^e_n$ from this Gaussian distribution and transform them into volumes $V^e_i=\frac{\pi (d^e_i)^3}{6}, i=1,\dots ,n.$ After normalization \eqref{vol}, we obtained the prescribed cell volumes throughout the window $Q_e$. According to the method in \cite{Bourne}, we compute unique (up to an additive constant) weights $w_i$, $i=1,\dots,n$, for the Laguerre tessellation $T_L$ in $Q_e$.

\subsubsection{Cell orientation distribution} 
The aim is to simulate the microstructure of a polycrystalline material with a cubic lattice orientation (such as, for example, in the B2 austenite phase of the NiTi alloy; see \cite{Alacron}) and to study the selected properties and implications of texture during deformation twinning. The texture is characterized by parameter $\kappa$ in the orientation distribution with odf \eqref{Preferential function}, assuming independent marking of the cells. Here, $\kappa=0$ yields a uniform distribution (i.e., no texture), while $\kappa=10,20,30$ represents increasing levels of texture. Moreover, for $\kappa =0$ we simulate another realization of the marked Laguerre tessellation with dependent marking using moving averages as defined in \eqref{ma}. The reference vector parameters $u=(0,0,1)^T$ and $v=(1,1,1)^T$ are kept fixed, $u$  coincides with the loading direction $D_l$, and $v$ corresponds to the main diagonal of the crystal lattice element in its reference coordinate system.

For the fixed loading direction $D_l=(0,0,1)^T$, the key quantities in this section, namely the Schmid factor $\chi$, the propensity for twinning $\Psi$, the normal direction $\vec{n}(G)$, and the volume fraction $V_t$ depend only on the orientation $G$ and can therefore be treated as marks of the tessellation cells. We consider $G$ as a random variable with probability density $f$ given in \eqref{Preferential function}, and investigate the sensitivity of the aforementioned quantities to the texture parameter $\kappa$.

\subsubsection{Presence of twinning}
In this study, the twinning system is $K_1 = (\overline{4}11)$, $K_2 = (011)$ (in Miller indices), and $\eta_1 = (1,2,2)^T$, $\eta_2 = (1,0,0)^T$; see \cite{Alacron}. Owing to the symmetries, we obtained twelve variants of these parameters.

Based on the propensity for twinning, we suggest a rule for twinning events in a given cell. Let $V_{max},\,V_{min}$ be the maximum and minimum cell volume among the cells that intersect $Q$, respectively. Note that thanks to plus sampling we can deal with all cells intersecting $Q$ because we know their total volume from $Q_e$. We choose two parameters $1>\psi_2>\psi_1>0,$ where $\psi_1$ is the critical propensity for the smallest cell and $\psi_2$ is the critical propensity for the largest cell. Thus, the critical propensity for twinning in a cell with volume $V$ is suggested to be 
\begin{equation}\label{prahy}\psi(V)=\psi_1+\frac{h(V)-h(V_{min})}{h(V_{max})-h(V_{min})}(\psi_2-\psi_1)\end{equation} for the function $h(V)=\big(\frac{6V}{\pi}\big)^{-1/6},\; V\geq 0.$  This depends on the cell volume according to the Hall--Petch relationship \cite{Hall,Petch}. 
The propensity for twinning \eqref{prope} is a function of the cell orientation $G$:
\begin{equation*} 
    \Psi =  \frac{1}{3 \sqrt{18}} \Big\langle \overline{R}(G)(-4,1,1)^T , {G} D_l \Big\rangle \Big\langle \overline{R}(G) (1,2,2)^T, {G} D_l \Big\rangle.
\end{equation*}
We assume that twinning occurs if and only if $\Psi\geq \psi(V)$. Thus, the probability of twinning depends on both propensity and cell volume.

\subsubsection{Simulation of lamellar systems} 
When twinning occurs in the cell $C$, we first generate an initial number $m$ of lamellae within $C$. To achieve this, we consider a random variable $M$ following a Poisson distribution that is right-truncated at $l_{max}-1$, with the parameter depending on $l_{max}$. We then set $m=M+1$. 
 
Next, the simulation of a random lamellar system in $C$ satisfying all conditions (i) to (iv) from Section \ref{sec:lamel} is carried out using one of the methods described therein.
 These methods require the entire cell to be fully observed, which is made possible through plus sampling. Having $C\subseteq Q_e$ intersecting $Q$, it is valid to realize the random lamellar system in the entire cell $C$ and its intersection with $Q$ provides the desired part of the lamellar system. The twinning normal vector \eqref{E1.33} is a function of cell orientation, and for the selected twinning system, it holds that
\begin{equation}
    \vec{n} = \frac{1}{\sqrt{18}}  G^{-1} \overline{R}(G)\begin{pmatrix}
        -4 \\ 1 \\ 1
    \end{pmatrix}.
\end{equation}
 
 \subsubsection{Meshing of nested tessellation and numerical simulation} \label{sec:4.2.5}
  The free software Neper \cite{Quey} is well-suited for modelling Laguerre tessellations. Although our nested tessellation $T_N$ with lamellae is no longer a Laguerre tessellation, Neper can still be used for storage and meshing. 
The finite element method (FEM), as described in \cite{MSCA}, was used to calculate the internal stresses. Each subcell of the tessellation $T_N$ within the window $Q$ was further subdivided into a fine tetrahedral mesh. Within each element of the mesh, the MSC Marc solver \cite{MSCA} provided approximations of the stress and inherent strain tensors. Tetrahedral elements of the ten-node quadratic (element 127 in \cite{MSCB}) were used. The total number of elements in the complete mesh depended on the number of lamellae. The stress tensor values are obtained at the nodes of the elements and then integrated to yield a single value for each element.

\subsubsection{Total strain energy density estimation} \label{sec:4.2.6}
Using the elements of the mesh, we obtain an estimate of the scalar total strain energy density \eqref{Tsed}:
\begin{equation}\label{tsed}
    \widehat{W} \;=\; \sum_{k=1}^{n_e} E_k \,|\mathcal{E}_k|,
\end{equation}
\noindent
with \begin{equation}E_k=\frac{1}{2}  \sum_{i=1}^3 \sum_{j=1}^3 \mathcal{E}_k^{\sigma_{ij}} \,\mathcal{E}_k^{\varepsilon_{ij}},\label{ek}\end{equation}
where \(\{\mathcal{E}_k : k = 1, \dots, n_e\}\) denotes the mesh of \([0,1]^3\),
and $\mathcal{E}_k^o$ are the extrapolated values of a function $o$ on $Q$ in the elements of the mesh. We set $o=\sigma_{ij}$ and $o=\varepsilon_{ij}$ in Eq. \eqref{ek}.

In addition, we divide $Q$ into two disjoint phases: the lamellar phase $L_P$, which is the union of all lamellar systems in cells, and the matrix phase $M_P$, where $Q=L_P\cup M_P.$ Their volume fractions in $Q$ are $V_L,\,V_M,$ respectively, where $V_L+V_M=1.$

Consequently, the tetrahedral mesh \( \{ \mathcal{E}_k : k = 1, \dots, n_e \} \) in $Q$, generated by Neper \cite{Quey}, allows us to partition the elements into two disjoint subsets:  
\begin{itemize}
    \item \( H_L \subseteq \{1, \dots, n_e\} \): indices of elements that lie entirely within $L_P$,
    \item \( H_M \subseteq \{1, \dots, n_e\} \): indices of elements that lie entirely within $M_P$.
\end{itemize}
Since $|Q|=1$, we have thus
\[
    V_L = \sum_{k \in H_L} |\mathcal{E}_k|.
\] 
Using this decomposition, the estimated total strain energy density in $L_P$ is defined as
\begin{equation} \label{lamspec}
    \widehat{W}_L = \frac{1}{ V_L} \sum_{k \in H_L} E_k \, |\mathcal{E}_k|,
\end{equation}
and in $M_P$ as
\begin{equation}\label{matspec}
    \widehat{W}_M = \frac{1}{V_M} \sum_{k \in H_M} E_k \, |\mathcal{E}_k|.
\end{equation}

This phase-specific energy decomposition enables us to examine how different microstructural regions contribute to the overall mechanical response. Using linear regression \cite{Seber}, we investigate whether texture or macroscopic strain disproportionately affects one phase more than the other.
 
\section{Numerical outputs} \label{sec:Numerical}
In this section, we present the statistical and numerical results obtained throughout the procedure of the stochastic and numerical simulations introduced in Section \ref{sec:stoch}, including the final regression analysis. After describing the generation of marked Laguerre tessellation, we display the results of the Schmid factor and the volume fraction of lamellae in Section \ref{sec:5.1}. Section \ref{sec:5.2} is devoted to the statistical analysis of the lamellar systems in cells. A detailed discussion of the regression analysis of the total strain energy density (TSED) on model parameters is provided in Section \ref{sec:5.3}. Regarding physical units, we assume that the edge length of our unit cube $Q$ is $100\,\mu$m, so the length unit is $10^{-4}\,{\rm m}$, while TSED is expressed in MPa.

\begin{figure}[t]
\centering
\subfloat[]{%
\resizebox*{6.5cm}{!}{\includegraphics{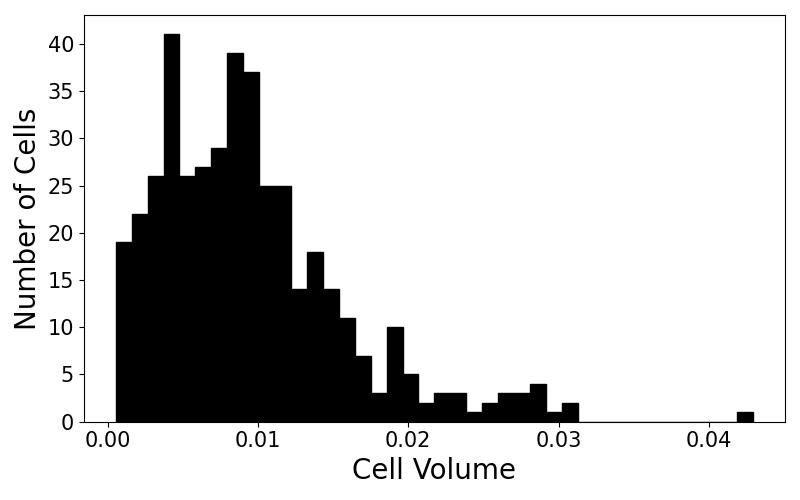}}}\hspace{10pt}
\subfloat[]{%
\resizebox*{6.5cm}{!}{\includegraphics{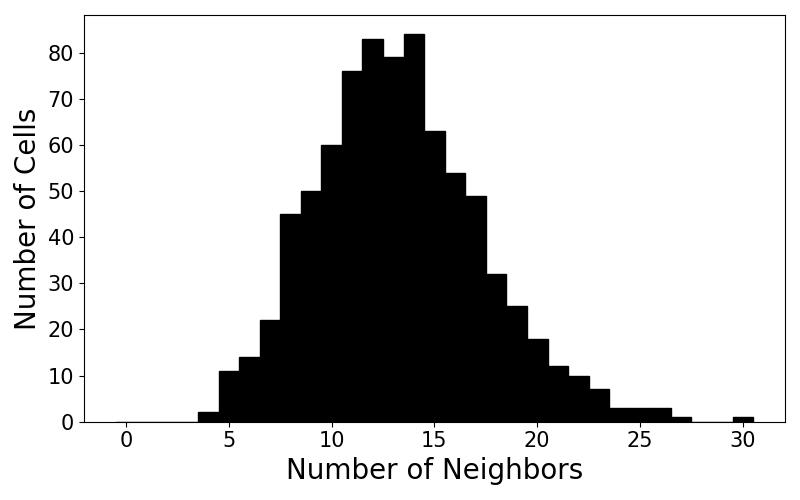}}}
\caption{Histogram of the cell volumes (a), histogram of the number of neighbors of inner cells (b). Sample size: $n=423.$}
\label{Figure: NM}
\end{figure}

First, we simulate a realization of the Laguerre tessellation $T_L$ according to Section \ref{sec:4.2.1} within the extended window $Q_e=[-0.5,\,1.5]^3$, using an intensity $\lambda_{0} =100$. In $Q_e$ we have 807 cells, including 423 inner cells (which do not intersect the boundary of $Q_e$). The number of cells that hit the observation window $Q=[0,1]^3$ is exactly 200.
In Figure \ref{Figure: NM}a we present the histogram of the volumes of the inner cells and in Figure \ref{Figure: NM}b the histogram of the number of neighbors of the inner cells. This realization of Laguerre tessellation was fixed for all subsequent analyses.
In the simulation of orientation distribution with odf \eqref{Preferential function}, we used four levels of texture parameters $\kappa =0,\,10,\,20,\,30$ with independent marking (IM) of the cells. 
For $\kappa =0$, we also considered the dependent marking by moving averages; see Eq. \eqref{ma}. The parameter $\varepsilon_m$ appearing in \eqref{mstrain} is taken on seven levels, from 0.05 to 0.2 with step 0.025. This means that we have 35 variants and for each of them a single realization of the marked Laguerre tessellation. Other parameters of the model are kept fixed, $l_{max}=3,\,\xi=0.05,\,\gamma=0.05,\,\zeta_1=0.05,\,\zeta_2=0.5,\,\theta=(1,\,1.2,\,1.2)^T.$

\subsection{Statistics of the existence and amount of twinning deformation} \label{sec:5.1}
In Eq. \eqref{prahy} of the critical propensity for twinning, we set $\psi_1=0.2$ and $\psi_2=0.4$ and have $V_{min}=0.0005$ and $V_{max}=0.043$. For $\kappa=0$ the relative frequency of twinning in a cell is equal to 0.97; in other cases, it is 1 (all cells undergo twinning). 

%

\begin{figure}[!ht]
    \centering
    \includegraphics[width=0.8\textwidth]{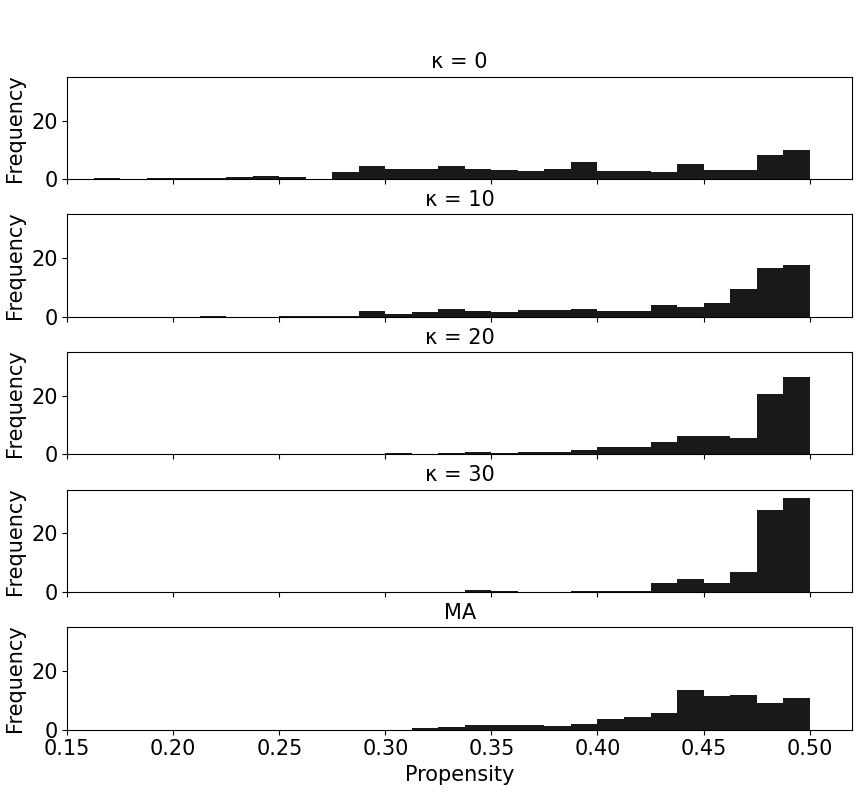}
    \caption{Histograms of the propensity for twinning are displayed for different values of $\kappa$ with IM and for the moving average model (MA). }
    \label{Figure: propensity}
\end{figure}
Figure \ref{Figure: propensity} shows that as $\kappa$ increases, the distribution of the propensity for twinning $\Psi $ shifts closer to the theoretical maximum of 0.5. The moving average case presented just a different orientation distribution with no texture; its propensity distribution is less skewed than in the cases with texture.

\begin{figure}[!ht]
    \centering
    \includegraphics[width=0.8\textwidth]{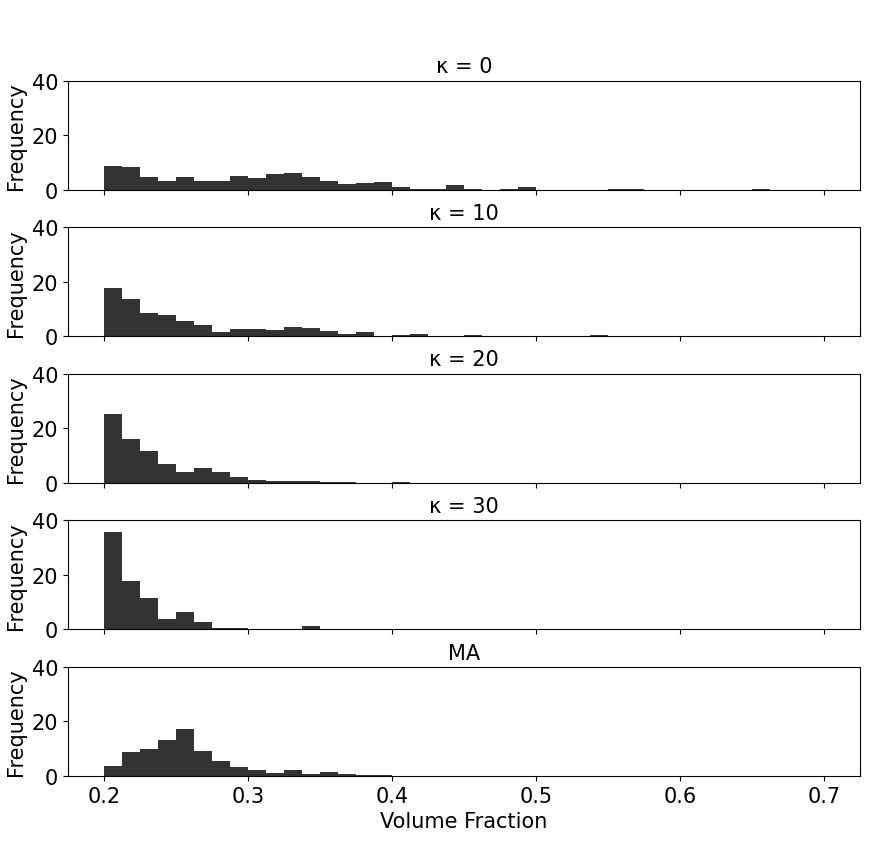}
    \caption{Histograms of the volume fraction $V_t$ of lamellae in cells are displayed for different values of $\kappa$ with IM and for the moving average model (MA).}
    \label{Figure: fraction}
\end{figure}

From (\ref{mstrain}), we can see that the volume fraction $V_t$ depends on the orientation of the crystal and the macroscopic strain $\varepsilon_m$. Because the latter serves only as a scaling factor, see Eq. \eqref{mstrain}, we set it to $0.1$ in Figure \ref{Figure: fraction}. In all five diagrams, the volume fraction of the lamellae exceeds 0.2; therefore, it follows that the strain induced by twinning satisfies $\overline{E}\leq \frac{1}{2}$ for any cell. In the histograms, we observe a stronger concentration of the volume fraction toward a value of 0.2 with increasing texture.
\begin{figure}[t]
\centering
\subfloat[Moving average]{%
\resizebox*{6cm}{!}{\includegraphics{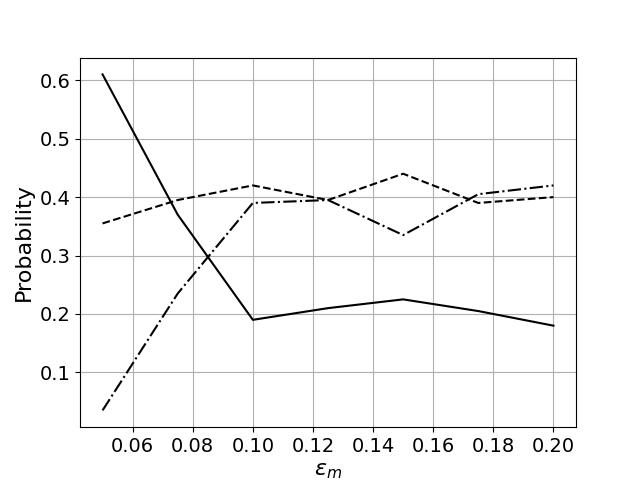}}}\hspace{5pt}
\subfloat[$\kappa=0$]{%
\resizebox*{6cm}{!}{\includegraphics{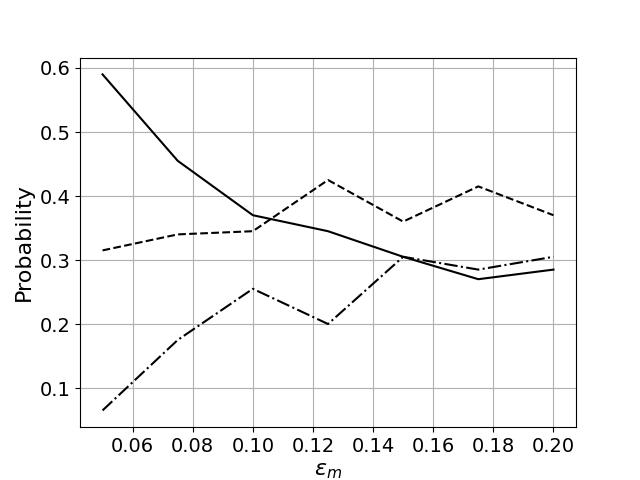}}}\vspace{5pt}
\subfloat[$\kappa=10$]{%
\resizebox*{6cm}{!}{\includegraphics{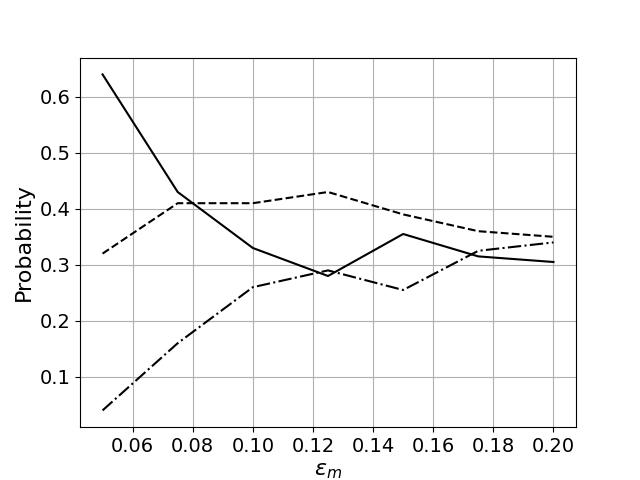}}}\hspace{5pt}
\subfloat[$\kappa=30$]{%
\resizebox*{6cm}{!}{\includegraphics{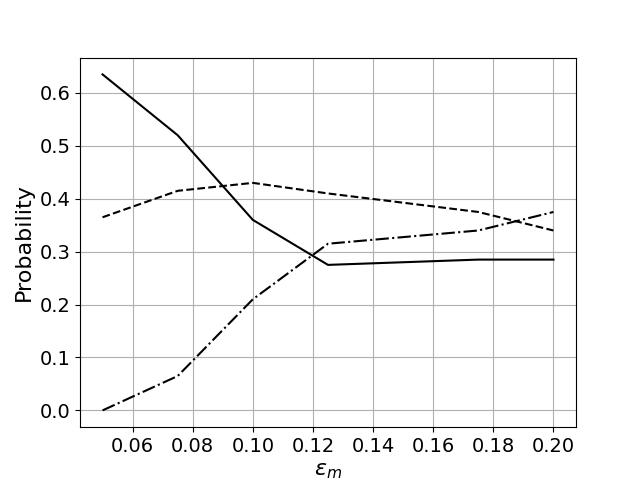}}}
\caption{Relative frequencies of lamellae counts in cells for $l_{max}=3$ with varying macroscopic strain levels \( \varepsilon_m \) under different marking conditions. Each subplot shows the evolution of the estimated probabilities (1 lamella -- solid lines, 2 lamellae -- dashed lines, 3 lamellae -- dash-dotted lines) with increasing \( \varepsilon_m \), for either the moving average or for independent marking with $\kappa=0,10,30$.}
    \label{fig:lamellae_probabilities}
\end{figure}

\subsection{Statistics of twin lamellae} \label{sec:5.2}
From the two methods of simulation of lamellar systems mentioned in Section \ref{sec:lamel}, we present the numerical results obtained using lamellar growth. This method is computationally faster, and the lamellar systems of both methods are observed to be similar.

First, we examine the number of lamellae per cell. This differs from the initial truncated Poisson distribution, because the results of the iterative lamellar growth depend on the macroscopic strain $\varepsilon_m$. The analysis is restricted to the inner cells $C_i$, $i=1,\dots,423$, of \( T_L \). We set $l_{max}=3$ and estimate the probabilities \( {p}_k \) of observing a cell with \( k = 1, 2, 3 \) lamellae by relative frequencies
\begin{equation*}
    \hat{p}_k = \frac{1}{423} \sum_{i=1}^{423}  \mathbb{I} \{ M_{i} = k \}  , \quad k =  1,2, 3,
\end{equation*}where $M_i$, $i=1,\dots,423$, are mutually independent random numbers of lamellae in the $i$-th cell.
In Figure \ref{fig:lamellae_probabilities} these estimates are plotted against the macroscopic strain $\varepsilon_m$ for different values of $\kappa$ under independent marking and moving average marking.  A consistent pattern naturally emerges across all random sampling models: as the macroscopic strain increases, the probability of observing a single lamella decreases, while the probability of two or three lamellae increases. Moreover, we observed that the probability of observing two lamellae remained approximately $0.4$ regardless of the value of $\kappa$ or the sampling scheme. 

\begin{figure}[h!]
\centering
\subfloat[$\kappa=0,\,\varepsilon_m=0.05$]{%
\resizebox*{9cm}{!}{\includegraphics{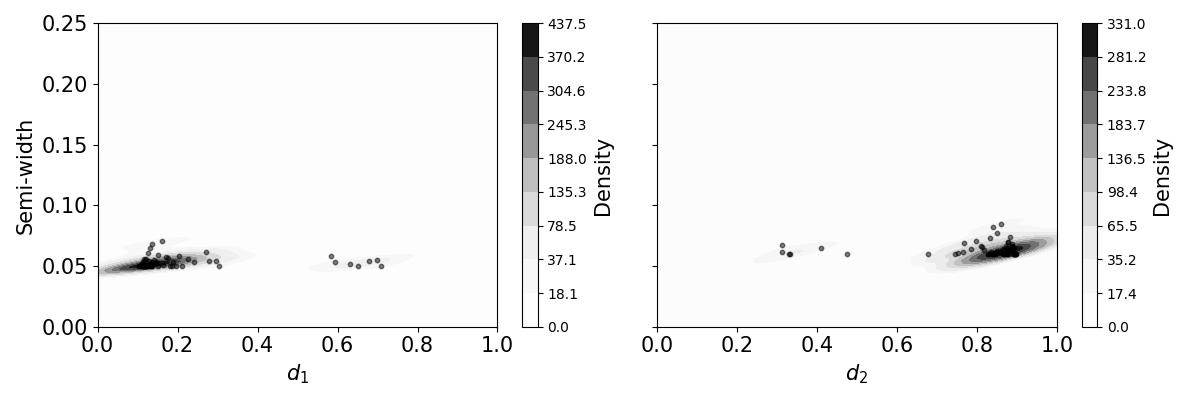}}}\vspace{5pt}
\subfloat[$\kappa=30,\,\varepsilon_m=0.05$]{%
\resizebox*{9cm}{!}{\includegraphics{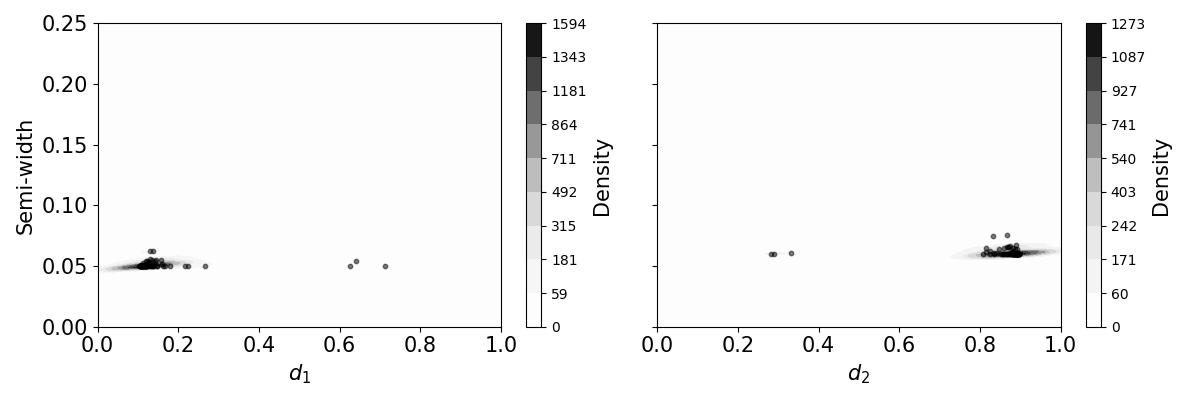}}}\vspace{5pt}
\subfloat[$\kappa=0,\,\varepsilon_m=0.2$]{%
\resizebox*{9cm}{!}{\includegraphics{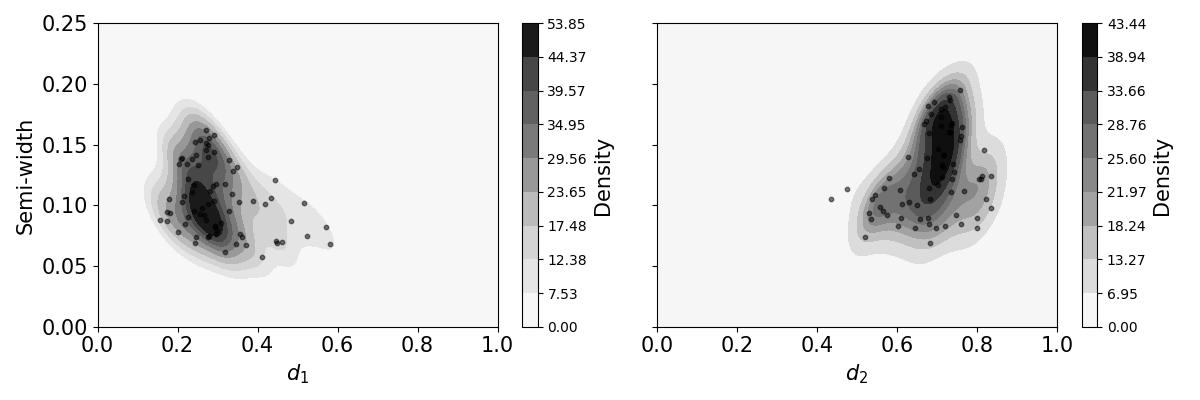}}}\vspace{5pt}
\subfloat[$\kappa=30,\,\varepsilon_m=0.2$]{%
\resizebox*{9cm}{!}{\includegraphics{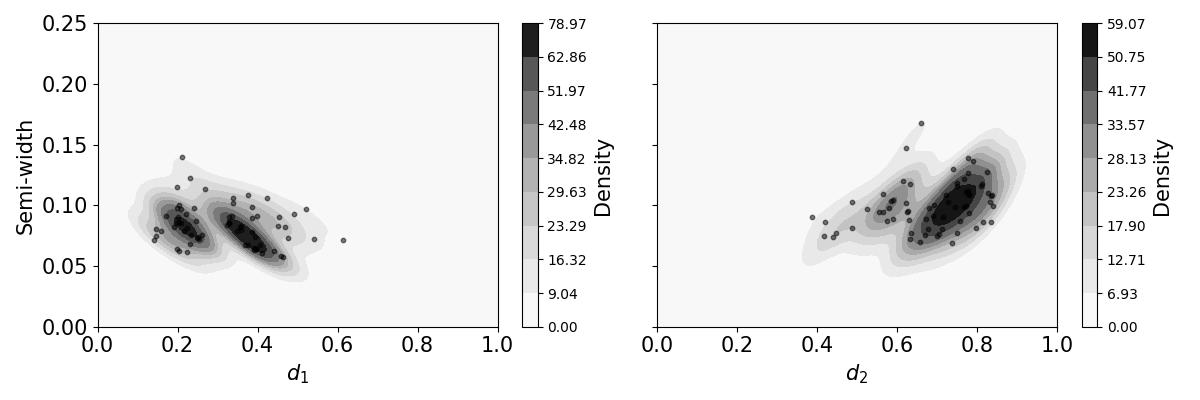}}}
\caption{2D kernel density estimates (KDEs) of normalized lamella center position and semi-width in cells with exactly two lamellae under different combinations of \( \kappa \) and \( \varepsilon_m \). The first, second lamellae are plotted on the left, right subfigure, respectively. KDEs are computed over lamella geometries normalized by the Feret diameter.}
\label{fig:kde_lamellae_center_width}
\end{figure}

To better understand the geometric characteristics of lamellae, we analyzed the distribution of their centers and semi-widths within the cells. We restricted this analysis to the configuration of cells containing exactly two lamellae. To account for variation in Feret projection lengths among cells, we normalized all lamellae geometries.  Let \( \widetilde{d}^{i}_k \) and \( \widetilde{w}^{i}_k \) denote the normalized center and semi-width of the \( k \)-th lamella in the \( i \)-th cell, with \( k = 1, 2 \) and \( i = 1, \dots, 423 \). These values were normalized using the Feret diameter $\rho^i$. As a result, normalized lamella centers lie in the interval \( (0,1) \), and normalized semi-widths lie in \( (0, \zeta_2/2) \).
We use two-dimensional kernel density estimation (KDE) to visualize the joint distribution of the normalized center position and semi-width.
We examine four representative cases corresponding to extreme values of \( \kappa \in \{0, 30\} \) and \( \varepsilon_m \in \{0.05, 0.2\} \) under independent marking. The KDE plots for these cases are shown in Figure~\ref{fig:kde_lamellae_center_width}.

  At low levels of macroscopic strain (\( \varepsilon_m = 0.05 \)), lamellae form predominantly near the ends of the Feret segment along the direction $\vec{n}$, as seen for both low and high \( \kappa \) (Figures~\ref{fig:kde_lamellae_center_width}a and~\ref{fig:kde_lamellae_center_width}b). This behavior indicates that only a small lamellar volume is required, which favors the peripheral placement. In addition, the semi-widths of the lamellae were smaller in this regime.
At high levels of strain (\( \varepsilon_m = 0.2 \)), lamellae tended to occupy more central regions of the cell (Figures~\ref{fig:kde_lamellae_center_width}c and~\ref{fig:kde_lamellae_center_width}d). This is consistent with the need to accommodate a larger lamellar volume and larger lamellar semi-widths.

\begin{figure}[!ht]
\centering
\subfloat[$\kappa=0,\,\varepsilon_m=0.05$]{%
\resizebox*{6cm}{!}{\includegraphics{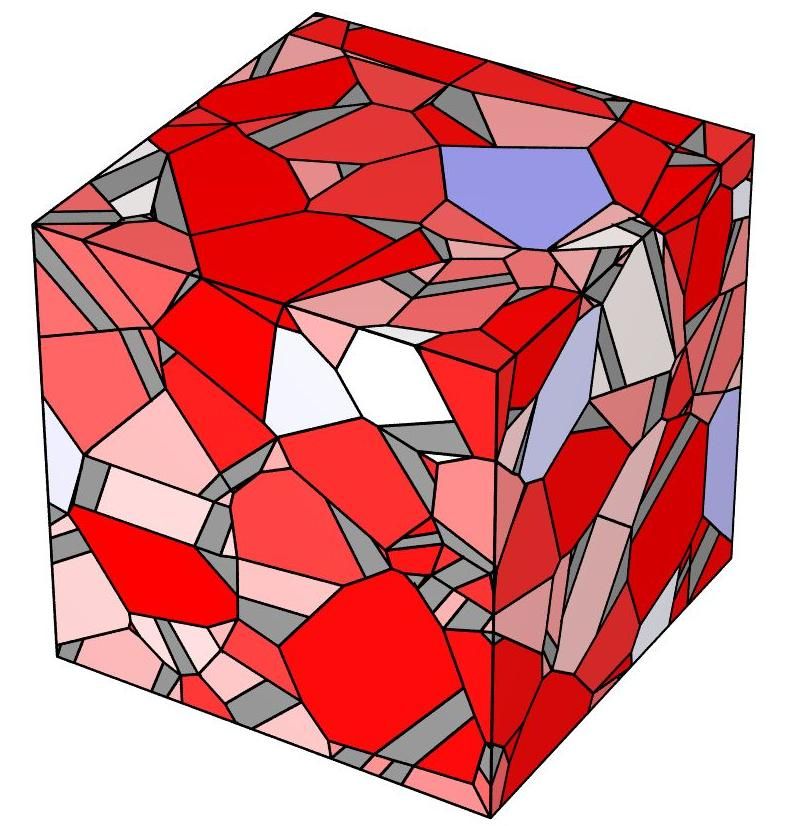}}}\hspace{5pt}
\subfloat[$\kappa=0,\,\varepsilon_m=0.2$]{%
\resizebox*{6cm}{!}{\includegraphics{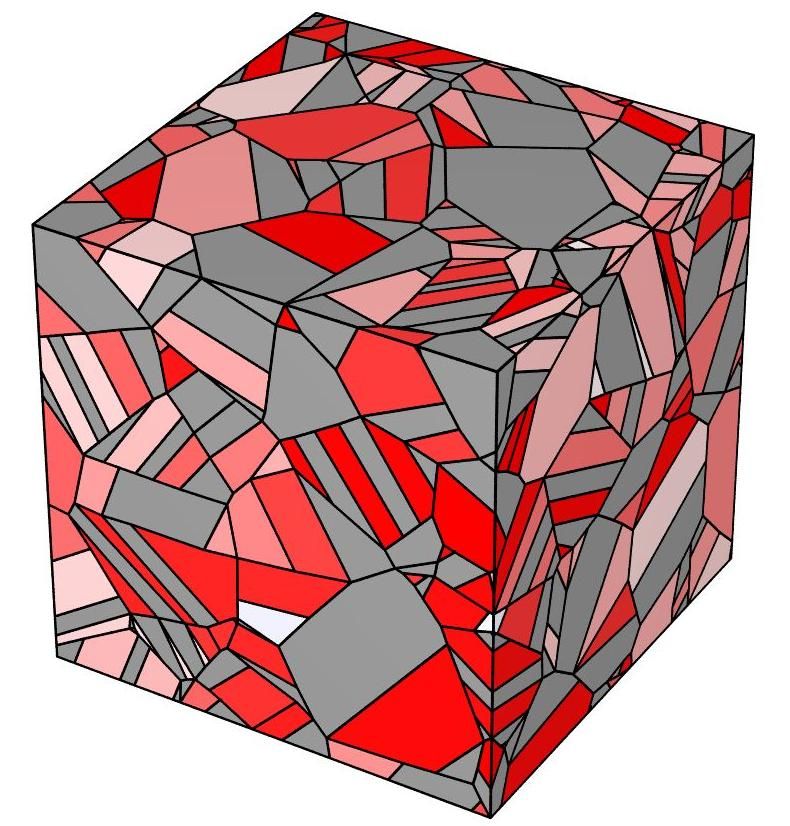}}}\vspace{5pt}
\subfloat[$\kappa=30,\,\varepsilon_m=0.05$]{%
\resizebox*{6cm}{!}{\includegraphics{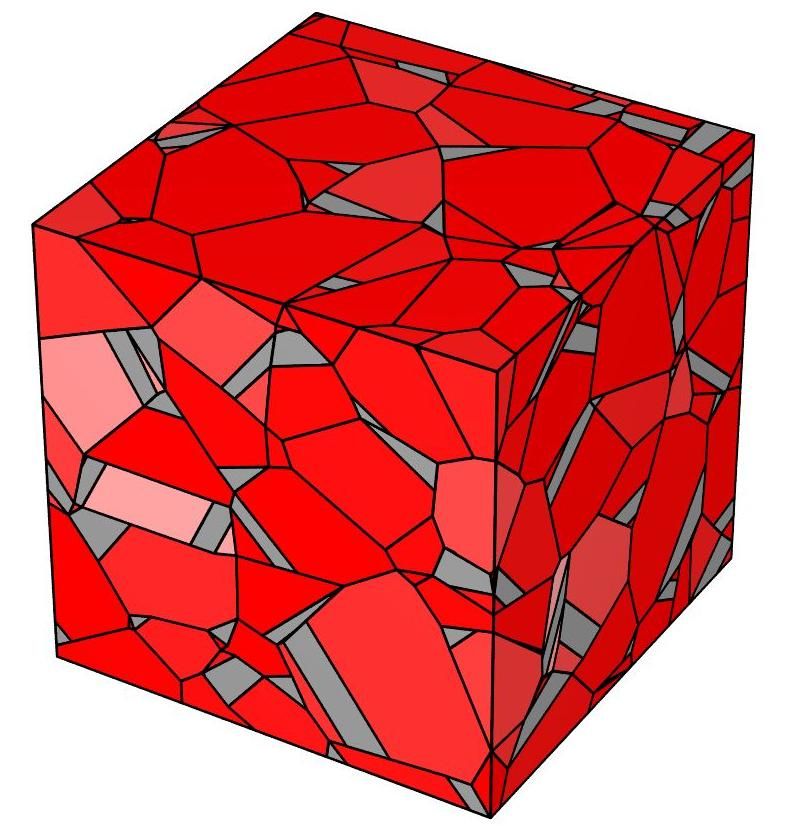}}}\hspace{5pt}
\subfloat[$\kappa=30,\,\varepsilon_m=0.2$]{%
\resizebox*{6cm}{!}{\includegraphics{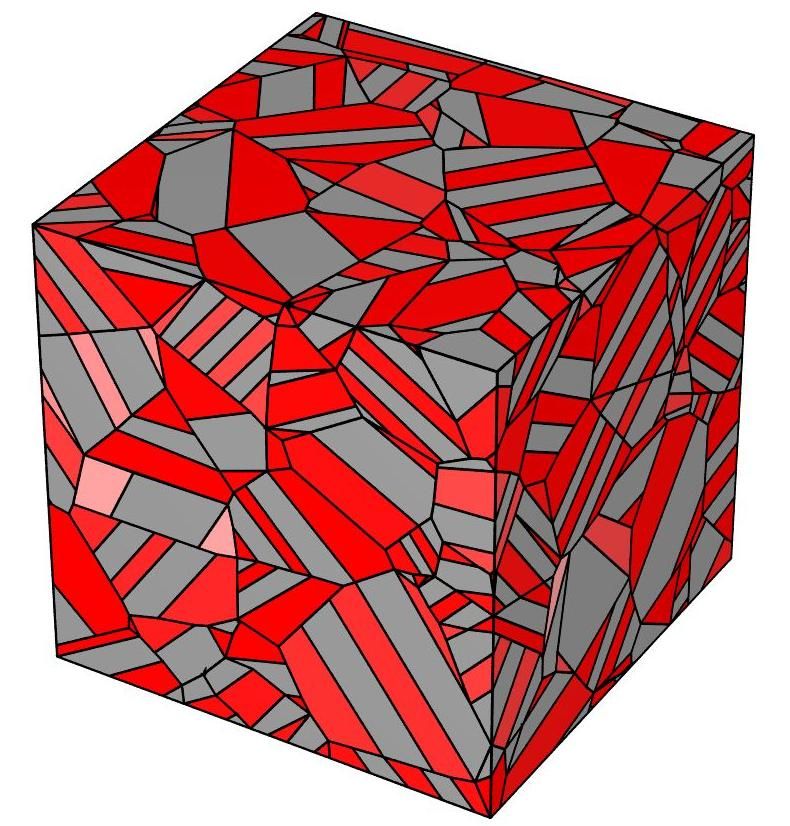}}}
\vspace{0.5cm}
\resizebox*{6cm}{!}{\includegraphics{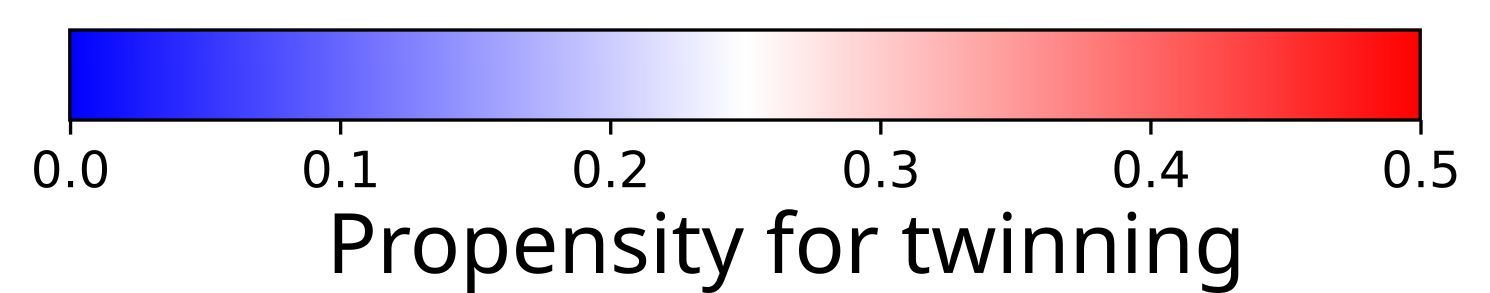}}
 \caption{Visualization of the marked nested tessellation $T_N$ in $Q$ for $l_{max}=3$ with cells colored according to the propensity for twinning. The subcells corresponding to lamellae are gray.} 
 \label{LamSchmid}
\end{figure}

In Figure \ref{LamSchmid} simulated marked nested tessellations are displayed where the lamellae are gray and the subcells of the matrix are colored according to the propensity for twinning of the corresponding mother cell. We observe three perpendicular planar boundaries of the 3D window $Q$, nevertheless the information on the placement of the lamellae and Figure \ref{Figure: propensity} is appropriately integrated. There are much fewer lamellae for $\varepsilon_m=0.05$ in subfigures a) and c). On the other hand, we observe a higher propensity for twinning in c) and d), which corresponds to the difference in histograms for $\kappa=0$ and $\kappa=30$ in Figure \ref{Figure: propensity}.

\begin{figure}[t]
\centering
\subfloat[]{%
\resizebox*{6cm}{!}{\includegraphics{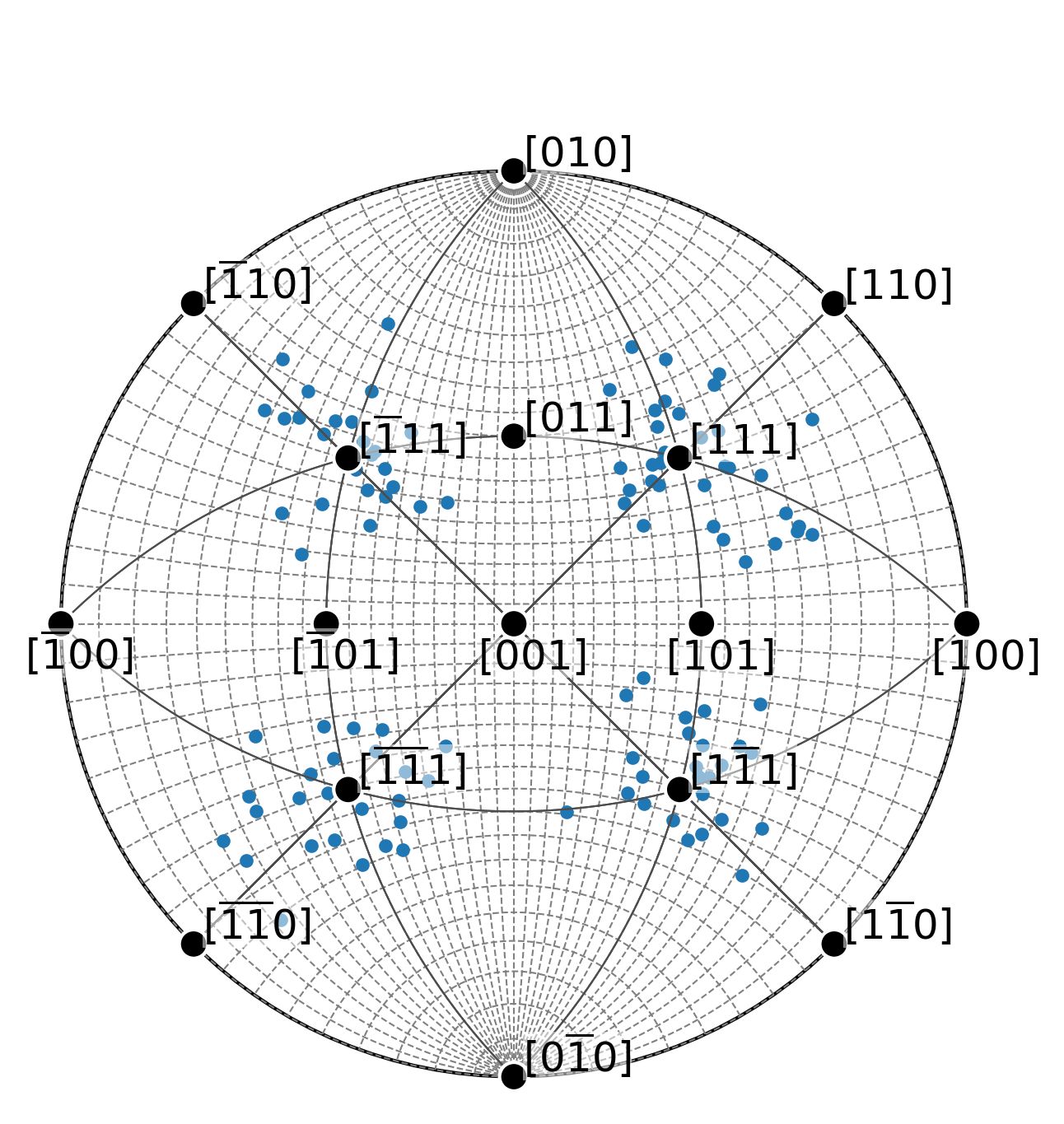}}}\hspace{10pt}
\subfloat[]{%
\resizebox*{6cm}{!}{\includegraphics{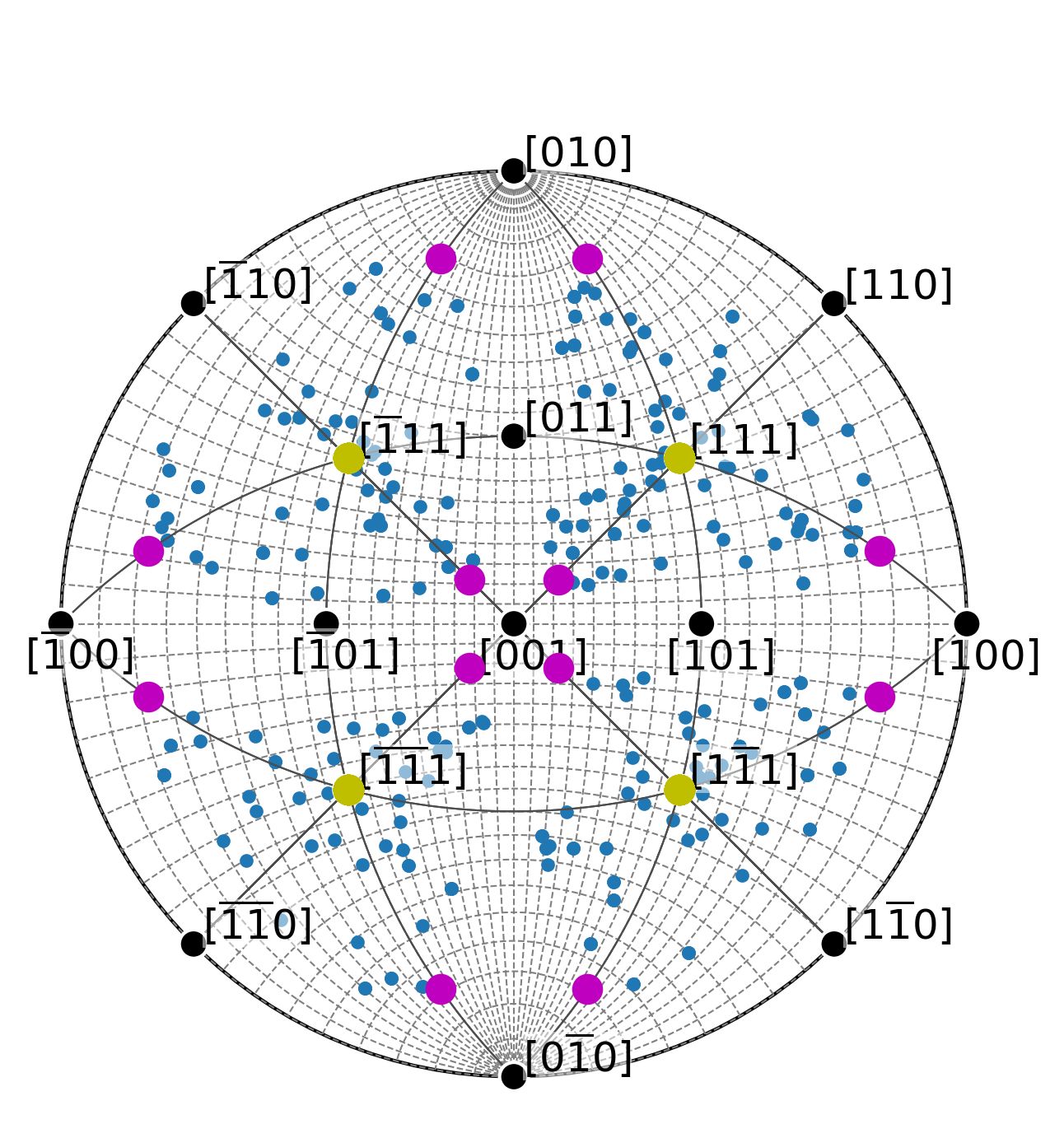}}}
\caption{(a) Inverse pole figure of $(0,0,1)^T$ direction for mother cells before twinning for the case of $\langle 111\rangle$ preferential orientation with $\kappa= 30$. (b) Corresponding inverse pole figure of $(0,0,1)^T$ direction for all subcells after twinning (blue poles) for the case of macroscopic strain of 0.2. In addition, the reorientation due to $\{114\}$ twinning is illustrated by four $\langle 111\rangle$ poles (yellow dots) that may reorient under tension into twelve orientations represented by magenta poles. The reorientation of four $\langle 111\rangle$ poles take place in $\{110\}$ planes denoted by thin black solid lines.   }
    \label{Figure: NN}
\end{figure}

Finally, we demonstrate the change in the orientation distribution during deformation twinning. We assume the case of $\langle 111\rangle$ texture with $\kappa =30$. This texture is characterized by an inverse pole figure of the $(0,0,1)^T$ direction, the poles of which are concentrated around the $\langle 111\rangle$ directions (Figure \ref{Figure: NN}a). After the deformation twinning simulated for macroscopic strain $\varepsilon_m=0.2$ the poles spread out from $\langle 111\rangle$ poles as shown in Figure \ref{Figure: NN}b. The orientation spread creates three clusters around each of the $\langle 111\rangle $ poles. The clusters are located near the $\{110\}$ planes, the traces of which are plotted in Figure \ref{Figure: NN}b by thin solid lines. This spread in orientation is a consequence of the lattice reorientation due to the $\{114\}$ twinning. This reorientation is characterized by rotations $R_t$, see \eqref{reor}, which occur in the $\{110\}$ planes. The reorientation due to these rotations is shown in Figure \ref{Figure: NN}b explicitly for four $\langle 111\rangle$ poles marked in yellow that may reorient under tension into twelve magenta poles (violet dots) that belong to one of the $\{110\}$ planes. Each of the four $\langle 111\rangle$ poles may reorient in three different orientations with the same probability, since the Schmid factors of the corresponding twinning systems are the same.    

\subsection{Linear regression analysis of model parameters} \label{sec:5.3}

The numerical simulation of the stress and strain tensors by FEM was computed as described in Section \ref{sec:4.2.5} for the 35 variants of marked nested tessellation. For each variant, we consider a single realization. Across these variants, the number of elements in $Q$ ranged from 337$\,$158 to 378$\,$068. We will index the parameters as follows: $$\kappa_j =10\,(j-1),\,j=1,\dots,4,\;\;\varepsilon_{m,i}=0.05+0.025\,(i-1),\,i=1,\dots ,7.$$
The main objective of this section is to examine how the total strain energy density (TSED), see \eqref{Tsed}, within the domain $Q=[0,1]^3$, is sensitive to changes in the above parameters governing the texture and macroscopic strain. In Section \ref{sec:5.3.1}, we present a detailed regression analysis; later, other cases are shortened to provide the final results.

\subsubsection{Total strain energy density under independent marking}
\label{sec:5.3.1}

For each combination of parameters $\kappa $ and $\varepsilon_m$, estimator \eqref{tsed} of the TSED was evaluated based on a single simulation.
We consider a linear regression model for these data in which the dependent variable is TSED and the independent variables are the parameters \( \kappa \) and \( \varepsilon_m \). The physical constraint that the TSED must vanish when \( \varepsilon_m = 0 \) must be satisfied. Thus, we define the following model:
\begin{equation} \label{model1}
    \widehat{W}_{i,j} = \beta_1 \varepsilon_{m,i} + \beta_2 \varepsilon_{m,i}^2 + \beta_3 \varepsilon_{m,i} \cdot \kappa_j + \beta_4 \varepsilon_{m,i}^2 \cdot \kappa_j + \eta_{i,j}, \quad j = 1, \dots,4,\; i=1,\dots, 7,
\end{equation}
where \( \eta_{i,j} \) is the error term.  

The first two coefficients, \( \beta_1 \) and \( \beta_2 \), capture the baseline effect (\( \kappa = 0 \)) of macroscopic strain on TSED. The interaction terms \( \beta_3 \) and \( \beta_4 \) quantify how texture (governed by \( \kappa \)) alters the linear and quadratic effects of \( \varepsilon_m \), respectively. To evaluate the model, we began by applying individual \( t \)-tests to each regression coefficient.
\begin{equation} \label{test3}
    H_0 : \ \beta_j = 0 \quad \text{versus} \quad H_1 : \ \beta_j \neq 0, \quad j = 1, \dots, 4.
\end{equation}
Subsequently, we performed a joint \( F \)-test to assess the significance of the interaction terms involving \( \kappa \):
\begin{equation} \label{test4}
    H_0 : \ \beta_3 = \beta_4 = 0 \quad \text{versus} \quad H_1 : \ \beta_3 \neq 0 \ \text{or} \ \beta_4 \neq 0,
\end{equation}
which tests whether the concentration parameter \( \kappa \) has a statistically significant effect on the influence of \( \varepsilon_m \) on the TSED.

Table~\ref{tab:random_kappa_model} presents the estimated regression coefficients
for \eqref{model1} and the results of the $t$-tests for hypotheses \eqref{test3}.
The results confirm that macroscopic deformation \( \varepsilon_m \) is the main driver of the TSED, with a strong positive linear effect and a significant negative quadratic term. 
Furthermore, we observe that the linear interaction term \( \beta_3 \), which models how the concentration parameter \( \kappa \) modulates the effect of \( \varepsilon_m \), is statistically significant (\( p = 0.0036 \)). In contrast, the quadratic interaction term \( \beta_4 \) is not significant (\( p = 0.7903 \)), suggesting that there was no appreciable nonlinear interaction between \( \varepsilon_m^2 \) and \( \kappa \).
The null hypothesis \eqref{test4} is rejected because the \( F \) statistic of 75.818 corresponds to a \( p \) value that is effectively equal to zero.

\begin{table}[h]
\caption{Regression results for model \eqref{model1}. Estimates of coefficients are shown with their standard errors, \( t \)-statistics, and \( p \)-values for the tests of the hypotheses \eqref{test3}. 
}
    \centering
    \begin{tabular}{|l|r|r|r|r|}
        \hline
        Coefficient & Estimate & Std. Error & \( t \)-value & \( p \)-value \\
        \hline
        \( \beta_1 \) & 1082.31 & 41.45 & 26.11 & 0.0000 \\
        \( \beta_2 \) & $-2270.73$ & 252.24 & $-9.00$ & 0.0000 \\
        \( \beta_3 \) & $-7.15$ & 2.22 & $-3.23$ & 0.0036 \\
        \( \beta_4 \) & 3.63 & 13.48 & 0.27 & 0.7903 \\
        \hline
    \end{tabular}
        \label{tab:random_kappa_model}
\end{table}

Based on the hypothesis testing results, we refined the model by removing the nonsignificant quadratic interaction term. The resulting reduced model, denoted by $M'_1$, is defined as
\begin{equation} \label{model1p}
    \widehat{W}_{i,j} = \beta_1 \varepsilon_{m,i} + \beta_2 \varepsilon_{m,i}^2 + \beta_3 \varepsilon_{m,i} \cdot \kappa_j + \eta_{i,j}, \quad j = 1, \dots, 4,\;i=1,\dots ,7.
\end{equation}

\begin{figure}
    \centering
    \includegraphics[width=0.7\textwidth]{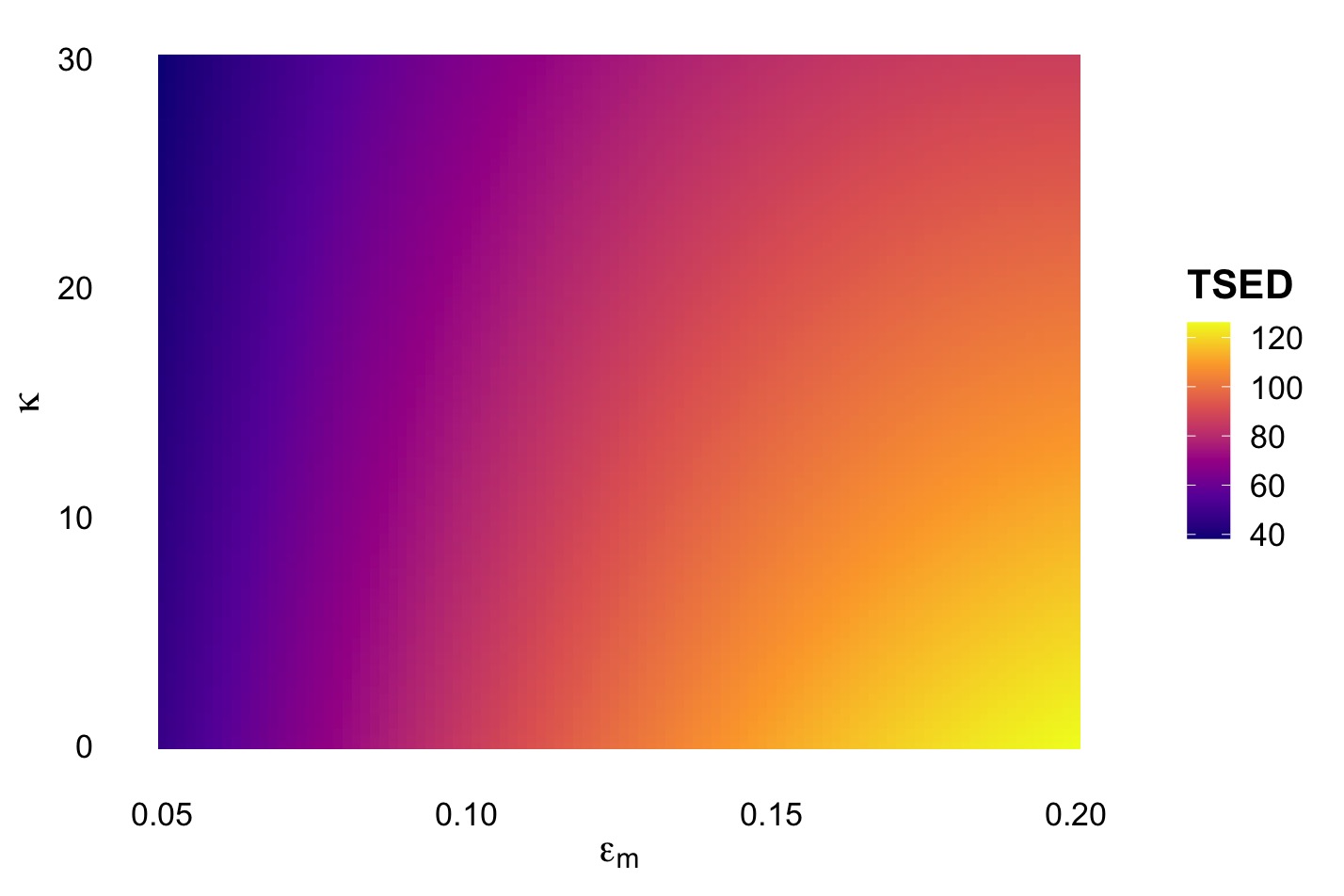}
    \caption{The fitted regression surface from model \eqref{model1p}, showing the joint effect of macroscopic strain \( \varepsilon_m \) and texture \( \kappa \) on TSED. Increasing \( \kappa \) leads to a reduction in the strain energy response, indicating enhanced energy dissipation in more aligned systems.}
    \label{fig:tsed_contour}
\end{figure}

The estimated coefficients for model \eqref{model1p} and the outputs of the $t$-tests are presented in Table~\ref{tab:regression_M2_prime}. To better understand the behavior of the model, we visualized the fitted surface as shown in Figure~\ref{fig:tsed_contour}. The plot clearly illustrates that higher values of the texture parameter \( \kappa \) attenuate the influence of macroscopic strain on the total strain energy density, indicating a more efficient distribution of internal energy under increased texture.

\begin{table}[h]
 \caption{Regression results for model \eqref{model1p}. Estimates of the coefficients are provided alongside their standard errors, \( t \)-statistics, and \( p \)-values for the tests of \eqref{test3}. 
 }
    \centering
    \begin{tabular}{|c|r|r|r|r|}
        \hline
        Coefficient & Estimate & Std. Error & \( t \)-value & \( p \)-value \\
        \hline
        \( \beta_1 \) & 1073.64 & 25.55 & 42.02 & 0.00 \\
        \( \beta_2 \) & $-2216.34$ & 147.92 & $-14.98$ & 0.00 \\
        \( \beta_3 \) & $-6.58$ & 0.52 & $-12.55$ & 0.00 \\
        \hline
    \end{tabular}
       \label{tab:regression_M2_prime}
\end{table}

\begin{figure}[t]
\centering
\subfloat[]{%
\resizebox*{5cm}{!}{\includegraphics{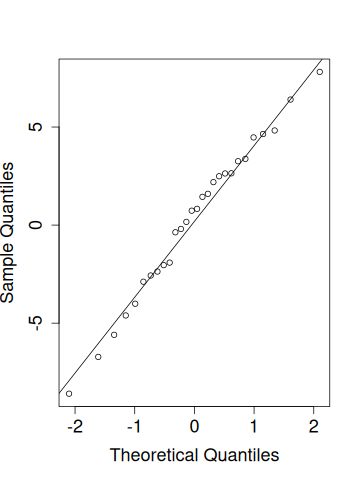}}}\hspace{10pt}
\subfloat[]{%
\resizebox*{5cm}{!}{\includegraphics{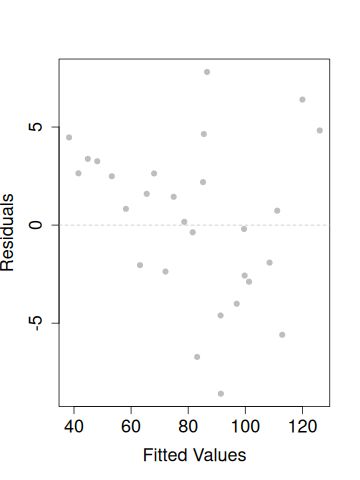}}}
\caption{Residual diagnostics for the regression model \eqref{model1p} analyzing total strain energy density (TSED) under independent marking. (a) Normal Q-Q plot of residuals, (b) residuals vs. fitted values.}
    \label{fig:TSED_random_diagnostics}
\end{figure}

Figure~\ref{fig:TSED_random_diagnostics} shows the diagnostic plots for the regression model \eqref{model1p}.  The normal Q-Q plot in Figure \ref{fig:TSED_random_diagnostics}a shows that the residuals are approximately normally distributed, with only slight tail deviations. The plot of residuals vs. fitted values in Figure \ref{fig:TSED_random_diagnostics}b does not show a discernible trend or funnel pattern, indicating a constant variance (homoscedasticity) in the range of fitted values.

\subsubsection{Separate total strain energy density analysis for lamellae and matrix}

To gain a deeper understanding of the internal energy distribution, we computed a separate regression analysis for the contributions to the strain energy density arising from the lamellae (phase $L_P$) and the matrix (phase $M_P$), as defined in Section \ref{sec:4.2.6}. We proceed analogously to Section \ref{sec:5.3.1}. Therefore, we restricted ourselves to a brief description of the results.

We begin our phase-specific analysis by focusing on the total strain energy density within the lamellae; see Eq. \eqref{lamspec}. In this context, our objective is to assess how the macroscopic strain $\varepsilon_m$ and  texture parameter $\kappa$  jointly influence the TSED stored in the lamellar regions.
To quantify these effects, we propose the following regression model,
\begin{equation}\label{M2}
   \widehat{W}_{L,i,j} = \beta_0 + \beta_1 \varepsilon_{m,i} + \beta_2 \varepsilon_{m,i}^2 
            + \beta_3 \kappa_j \varepsilon_{m,i} + \beta_4 \kappa_j \varepsilon_{m,i}^2 + \eta_{i,j},
\end{equation}
for $j = 1, \dots, 4$, $i=1,\dots ,7$.
Here, \( \widehat{W}_{L,i,j} \) denotes the estimated lamellae-specific TSED, see Eq. \eqref{lamspec}, for the \((i, j) \)-th variant, and \( \eta_{i,j} \) represents the error term.

\begin{figure}[t]
\centering
\subfloat[]{%
\resizebox*{7cm}{!}
{\includegraphics{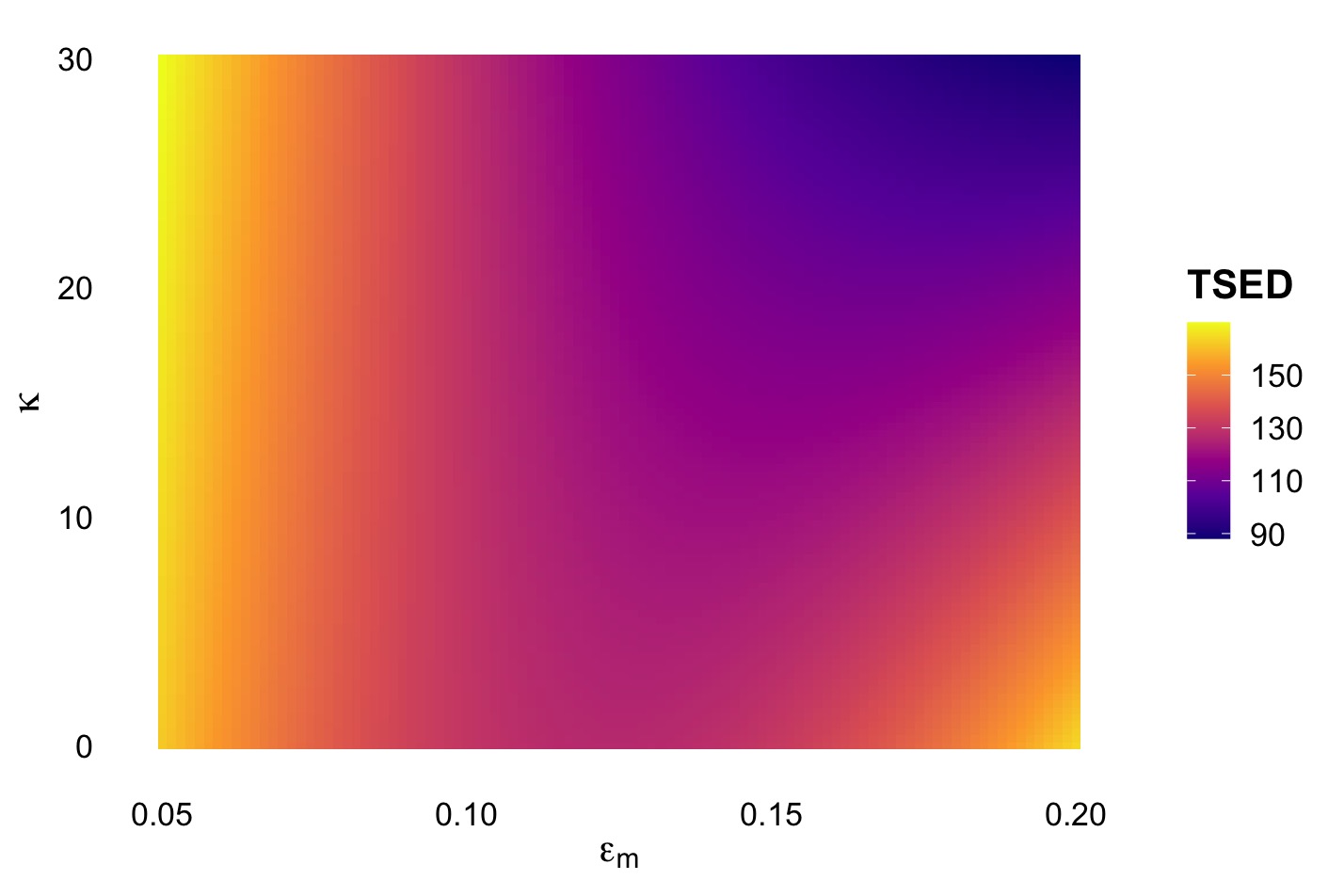}}}\vspace{5pt}
 \subfloat[]{%
\resizebox*{7cm}{!}{\includegraphics{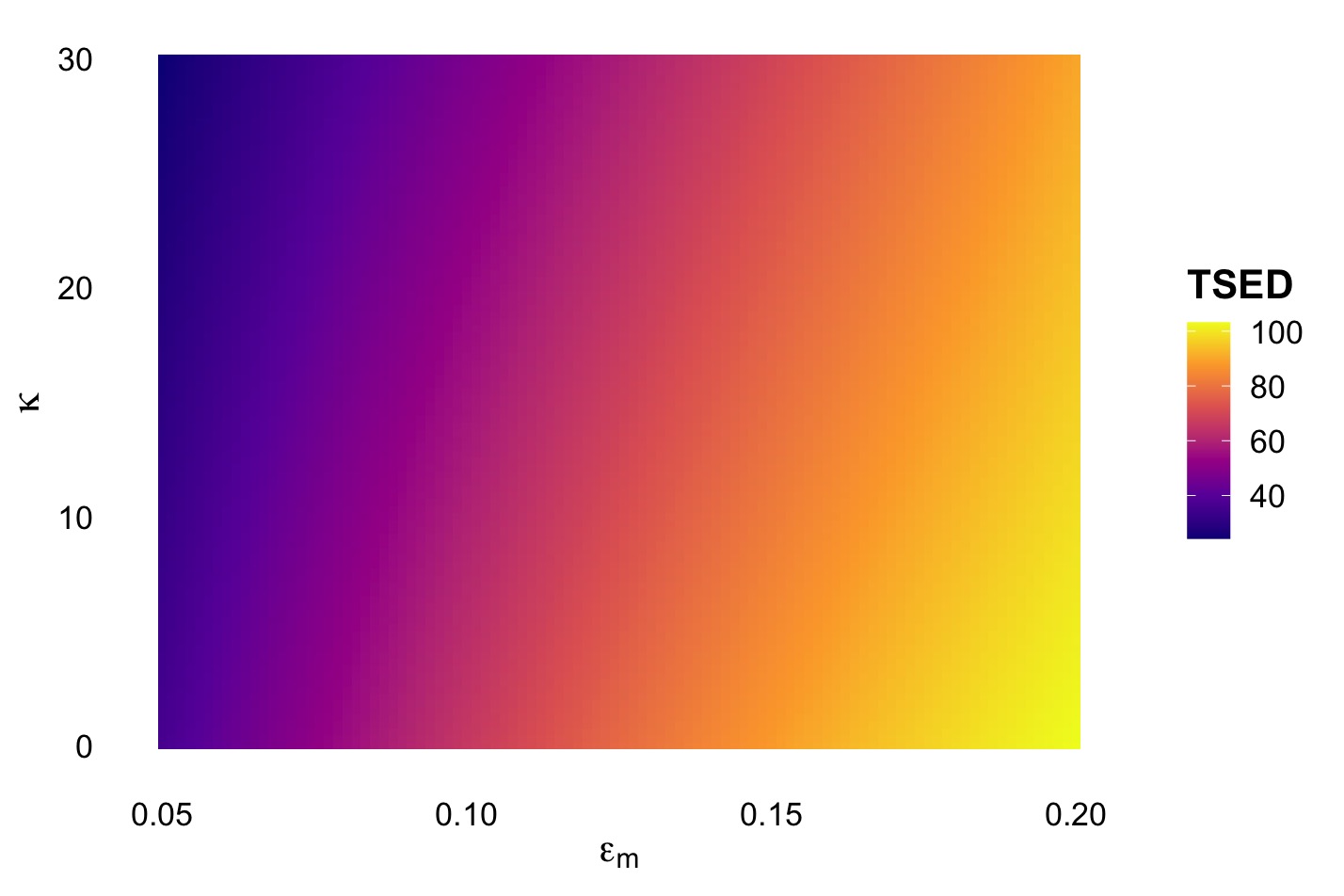}}}
\caption{Fitted regression surfaces of TSED as a function of macroscopic strain and texture parameter. (a) Model \eqref{M2} for the lamellar phase, (b) model \eqref{M3} 
for the matrix phase.}
\label{fig:matrix_contour}
    \end{figure}

The least-squares estimates are $\hat{\beta}_0=225.2$, $\hat{\beta}_1=-1590.9$, $\hat{\beta}_2=6420.8$, $\hat{\beta}_3=11.5$, $\hat{\beta}_4=-120.5$, all parameters are statistically significant. 
Figure \ref{fig:matrix_contour}a shows the fitted regression surface.
There is particularly strong evidence for both the quadratic strain term \( \beta_2 \) and the interaction terms involving the parameter \( \kappa \).
In particular, the positive value of \( \beta_3 \) and the negative value of \( \beta_4 \) suggest a nuanced interaction. Although increased texture mildly amplifies the linear effect of macroscopic strain on the lamellar energy, it simultaneously suppresses the quadratic contribution. 

Next, we analyze the total strain energy density in the matrix region; see Eq. \eqref{matspec}. We started with the same regression model as in \eqref{M2}; however, here the parameter $\beta_0$ is not significant. Therefore, we refine the regression by removing the intercept term. The reduced model
is defined as follows:
\begin{equation} \label{M3}
    \widehat{W}_{M,i,j} = \beta_1 \varepsilon_{m,i} + \beta_2 \varepsilon_{m,i}^2 + \beta_3 \kappa_j \varepsilon_{m,i} + \beta_4 \kappa_j \varepsilon_{m,i}^2 + \eta_{i,j},
\end{equation}
for \(i=1,\dots, 7,\, j = 1, \dots, 4 \). The least squares estimates are $\hat{\beta}_1=799.7$, $\hat{\beta}_2=-1419.0$, $\hat{\beta}_3=-10.1$, $\hat{\beta}_4=39.3$, and all coefficients are statistically significant.
The relationship between the TSED in the matrix, macroscopic strain, and texture was highly regular and well captured by the model. The fitted surface shown in Figure~\ref{fig:matrix_contour}b illustrates that the energy stored in the matrix increases with macroscopic strain, as expected. However, a higher \( \kappa \) consistently reduced the TSED in the phase $M_P$.

\subsubsection{Comparison of moving average and independent marking}
In this analysis, we examined how the method of marking tessellation cells by orientation affects the total strain energy density under varying levels of macroscopic strain. As the moving average model is defined only for $\kappa = 0$, we restrict our attention to this case. 
As before, the TSED vanishes when the macroscopic strain \( \varepsilon_m \) is zero. After evaluating several regression models, we identified the cubic model with a shared cubic term as the most appropriate. For independent marking, the model is expressed as follows
\begin{equation*}
   \widehat{W}_i = \beta_1 \varepsilon_{m,i} + \beta_2 \varepsilon_{m,i}^2 + \beta_5 \varepsilon_{m,i}^3 + \eta_i,\quad i=1,\dots,7,
\end{equation*}
while that for the moving averages is
\begin{equation*}
    \widehat{W}_i = \beta_3 \varepsilon_{m,i} + \beta_4 \varepsilon_{m,i}^2 + \beta_5 \varepsilon_{m,i}^3 +\eta_i, \quad i=1,\dots,7.
\end{equation*}
The least squares estimates are $\hat{\beta}_1=1324$, $\hat{\beta}_2=-5806$, $\hat{\beta}_3=1223$, $\hat{\beta}_4=-6370$, $\hat{\beta}_5=12727$, all coefficients are statistically significant. 
   
\begin{figure}[h]
    \centering
    \includegraphics[width=0.8\textwidth]{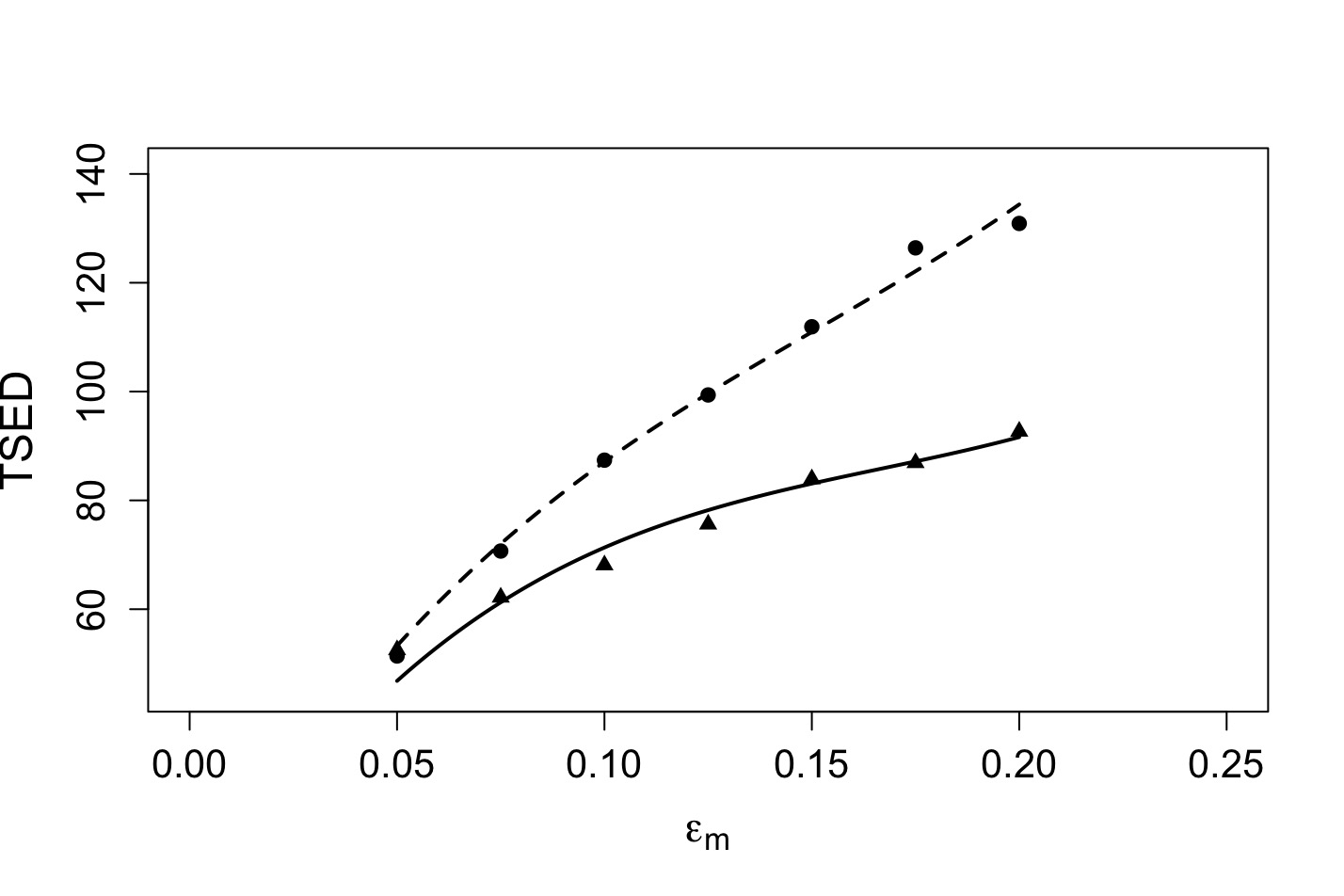}
    \caption{Visualization of the effect of macroscopic strain \( \varepsilon_m \) on TSED for the two marking models. Data points are indicated by dots (IM model) and triangles (MA model), respectively, with their corresponding fitted regression curves shown as dashed (IM) and solid (MA) lines.}
    \label{Figure: regression1}
\end{figure}

Figure~\ref{Figure: regression1} compares the relationship between the macroscopic strain \( \varepsilon_m \) and the TSED for the two marking models, based on the fitted regression curves. The results show that across the strain range, the IM model consistently yielded higher TSED values, with a sharper increase at higher macroscopic strains. In contrast, the MA model resulted in a moderate TSED response. These findings highlight the impact of the marking strategy on the predicted mechanical behavior.

\medskip
\section{Discussion}
The paper emerged from the cooperation between mathematicians (probabilists) and physicists, addressing a problem in materials science. Stochastic modelling of the microstructure of polycrystalline materials was previously developed step-by-step, starting with random Laguerre tessellation \cite{Petrich,Seitl}, and some authors \cite{Teferra,Sedivy} generalized this modelling to tessellations with nonplanar surfaces. The next step was to mark the tessellations by crystallographic orientations \cite{Pawlas,Karafiatova}. The most recent step is to obtain a finer scale, namely, on a subcell level of crystal defects. In \cite{Rieder} stochastic models, the twinned polycrystalline material was explored based on experimental data. 

In the present study, the rigorous theory of deformation twinning \cite[Chapter 11]{Kelly} was applied and the virtual microstructures of twinned polycrystals were simulated. We condition the realization of the Laguerre tessellation with a prescribed cell volume distribution. Our focus is on the variability of crystallographic orientations (called texture in physics), ranging from uniform random to preferential with a strong concentration around a mode. With such an increase in texture, a statistically increasing propensity for twinning was observed (\ref{Figure: propensity}). We also considered two types of orientation markings of cells: independent random or dependent, based on moving averages. The dependence leads to smaller disorientation angles, see \eqref{eq:dis}, of neighboring cells. This has no effect on the use of the Schmid factor when deciding about twin presence, since the propensity for twinning depends on orientation rather than disorientation angle, it is applied for each cell separately independently of the other cells.

Twinning is based on a parametric stochastic model with several geometric constraints. This model is empirical, and we tried two methods of simulating twin lamellae, which led to similar results in the distribution of lamellae locations and widths. Therefore, the numerical results for the method called 'lamellar growth' are presented. Although the model lacks analytical tractability, it is compatible with the meshing capabilities of Neper software \cite{Quey}, which can process the resulting nested tessellation, even though it is no longer of the Laguerre type. This enables the continuation of the stochastic simulation study with numerical simulations of internal stress and strain fields using the MSC Marc software \cite{MSCA}. We restricted ourselves to the total strain energy density (TSED) in scalar quantities in the final regression analysis.

The second main parameter investigated in the model is the macroscopic strain, which is directly proportional to the volume fraction of lamellae within the cells. When this parameter was increased, the total strain energy density increased. However, we observed a difference between the dependent and independent orientation markings (Figure \ref{Figure: regression1}), with the former leading to lower TSED values. The joint dependence of the TSED on both texture and macroscopic strain is shown in Figure \ref{fig:tsed_contour}. Next, this quantity is divided into two phases, lamellae and the matrix, and we observe a different form of this joint dependence in Figure \ref{fig:matrix_contour}a, where the TSED values are higher than those in Figure \ref{fig:tsed_contour}.

\section{Conclusions}
A new parametric stochastic model of a marked nested Laguerre tessellation was introduced, where the original marks of the mother cells corresponded to random crystallographic orientations. The formation of subcells is based on the theory of deformation twinning in polycrystalline materials under loading. The subcells were of two types: twin lamellae with a determined change in the mark and interlamellar spaces. Although the decision on the existence of twinning in a cell follows from physical theory, the formation of the lamellar system must be described by another stochastic model. Thus, we can address this complex model using stochastic simulations. An application presented is the numerical simulation of the internal stress and inherent strain on the marked nested tessellation, which is based on the elastoplastic behavior of the material. We conclude that the local smoothing of the mother cell orientations without texture leads to a decrease in the total strain energy density. When the lamellae and matrix phases are separated, the strain energy density in the matrix presents a trend with respect to the texture and macroscopic strain, which is similar to that of the TSED. However, the behavior of the lamellae is more irregular, which requires further investigation. The procedure developed in this study is valuable for further research on inhomogeneous internal stresses arising from deformation twinning in polycrystalline alloys.

\section*{Acknowledgments} This work was supported by the Czech Science Foundation under Grant 22-15763S; and the Ministry of Education, Youth, and Sports of the Czech Republic and European Union under Grant CZ.02.01.01/00/22\_008/0004591.

\end{document}